\documentstyle[psfig]{l-aa}     

\begin{document}
    \thesaurus{11;       
               11.13.1;  
               11.19.4;  
               11.19.5;  
               08.05.2;  
               08.05.1;  
               08.06.3  
             }

\title{Be star surveys with CCD photometry.
}
\subtitle{II. 
NGC 1818 and its neighbouring cluster in the LMC
\thanks{Based on observations taken at the European Southern Observatory,
La Silla, Chile}
}

\author{
Eva K.\ Grebel \inst{1,} \inst{2,} \inst{3}
}

\offprints{E.K.\ Grebel, AI W\"urzburg}
\institute{
Sternwarte der Universit\"at Bonn, Auf dem H\"ugel 71, D-53121 Bonn,
Germany
\and
University of Illinois at Urbana-Champaign, Department of Astronomy,
1002 West Green Street, Urbana, IL 61801, USA
\and
Astronomisches Institut der Universit\"at W\"urzburg, Am Hubland,
D-97074 W\"urzburg, Germany\\ grebel@astro.uni-wuerzburg.de
}

\date{Received Mar 19, 1996; accepted June 7, 1996}

\maketitle
\markboth{E.K.\ Grebel: Be star surveys with CCD photometry. II. NGC 1818}{}

\begin{abstract}
As part of an ongoing photometric survey of young Magellanic Cloud clusters
we identified Be stars in NGC 1818 and a nearby smaller cluster  
in the Large
Magellanic Cloud. The neighbouring cluster does not appear to contain any
evolved stars, and its sparsely populated main sequence does not extend to
stars as massive as in NGC 1818. Both clusters are younger than the 
surrounding field population, but the current data do not allow to conclude
whether NGC 1818 is a binary cluster or not. The small cluster is more 
heavily reddened than NGC 1818 indicating the presence of differential
reddening, leftover gas and dust from the star formation process, or a 
larger distance. NGC 1818 does not seem to be 
significantly affected by differential reddening. 

We find both clusters to be rich in Be stars\footnote
{Tab.\ 2, which contains identifiers, magnitudes, and coordinates of the Be
stars in NGC 1818 and surroundings will be made available only in
electronic form at the CDS via anonymous ftp to cdsarc.u-strasbg.fr
(130.79.128.5) or via http://cdsweb.u-strasbg.fr/Abstract.html.}.
The field only contains very few Be stars as one would expect for a
predominantly older population. NGC 1818
contains almost as many Be stars as the slightly younger SMC cluster NGC 330,
while NGC 2004, a young LMC cluster, has a lower Be star content.

We discuss problems in comparing Be star fractions in Magellanic Cloud
clusters and Galactic open clusters and possible constraints on Be star
theories. 

\end{abstract}

\section{Introduction}\label{sect_intro}

Be stars are non-supergiant (luminosity class V to III) B type stars 
that show or once have shown Balmer emission (Jaschek et al.\ 1981) -- 
a very wide definition. Among Galactic field B stars, 18\% of B0 to B7 
III--V stars are Be stars (Abt 1987).  While Galactic
Be stars have been extensively investigated, little is known about
the Be star content of the Magellanic Clouds.
The Magellanic Clouds have lower metallicity than usual for the young
stellar population in the Milky Way, which affects opacity and stellar
atmosphere properties, hence the study of Be stars in the
Magellanic Clouds may shed light on the formation mechanisms of the Be
phenomenon. The study of Be stars in
clusters has the added advantage of looking at a population 
homogeneous in age and metallicity, and of common origin.
Therefore, we have started a survey of Be stars in Magellanic Cloud 
clusters.
 
In the first paper of this Be star survey (Grebel et al.\ 1992: Paper I),
we studied the young cluster NGC 330 in the Small Magellanic Cloud (SMC)
and found it to be unusually rich in Be stars. Our photometric survey 
technique uses imaging in three filters, one of them an $H\alpha$ filter. 
The colour index formed by the two other filters serves to separate blue 
stars from red stars. A colour index formed from the $H\alpha$ filter and 
one of the other filters is used to distinguish stars 
bright in H$\alpha$ from the rest. In the resulting two-colour diagrams
(TCDs) Be star candidates stand out clearly. If the $H\alpha$ filter
is used together with a continuum filter such as Johnson-Cousins $R$,
the strength of the  H$\alpha$ emission can be calibrated
(Grebel et al.\ 1994). 

A comparison of the detections from the photometric survey of NGC 330 with
those from grism surveys shows excellent agreement (Paper I). The photometric
method has the added advantage of being able to operate also in very crowded 
regions and of allowing to distinguish between Be stars and ordinary B stars.  
Medium-resolution spectroscopy (Lennon et al.\ 1994, Grebel et al.\
1996) allows one to detect very weak-lined Balmer emission stars that 
can not be distinguished by photometric techniques, though in practice 
medium-resolution spectroscopy is restricted to few stars with the currently 
available instruments owing to constraints on observing time.  CCD imaging 
with two broadband filters and an $H\alpha$ filter is an efficient tool to 
identify Be stars.  The definition of a Be 
star implies that {\it all current detections establish only a lower limit 
to the true number of Be stars}. 

In the present paper, the Be star content of NGC 1818 and the neighbouring
smaller cluster in the Large Magellanic Cloud (LMC) will be analyzed (Sect.\ 
\ref{sect_n1818be}). We compare Be star fractions in young
clusters in the Magellanic Clouds and in the Milky Way and discuss the 
associated problems in Sect.\ \ref{sectbe_prob}.  In Sect.\ 
\ref{sectbe_constr}  we discuss constraints on Be star theories.

\section{The photometric data}\label{sect_photdata}

$H\alpha$ frames of NGC 1818 were obtained at the New Technology 
Telescope (NTT) with the ESO Multi Mode Instrument (EMMI) on 03 Feb 1992
at ESO, La Silla, Chile. A Thomson $1024^2$ chip (ESO \# 18) with a
pixel scale of $0\farcs43$ resulting in a field of view of $7\farcm3 \times
7\farcm3$ was used. The $H\alpha$ filter was \# 654 ($\lambda_c = 6554$ \AA,
$\Delta\lambda = 33$ \AA). 

We observed one field centered on NGC 1818 on 08 Oct 1993
with the 2.2m MPIA telescope at ESO, La Silla, Chile using
the ESO Faint Object Spectrograph and Camera (EFOSC2) with a 1024 
$\times$ 1024 Thomson chip (ESO \# 19). The pixel scale was $0\farcs332$, 
which gives a field of view of $5\farcm7 \times 5\farcm7$.
We obtained short and long exposures in Bessell {\em UBVR}.
The $R$ filter was ESO \#585 (Bessell $R$, $\lambda_c = 5949$ \AA,
$\Delta\lambda = 1654$ \AA) from the standard filter set for the 2.2m
telescope. We also observed several {\em UBV(RI)}$_{\rm C}$ standard 
fields from Landolt (1992) in the same night resulting in a total of 47 
standard star observations. The data were reduced using the standard 
procedures in DAOPHOT II running under MIDAS (Stetson 1992).
As turned out later, during part of 1993 the telescope was mistakenly 
operated without sky baffle. Our $U$ and $B$ frames (and to a lesser
extent, the $V$ frame) show systematic position-dependent magnitude
variations of up to $\pm 0.2$ magnitudes. The effect on the $R$ filter
is negligible. Such a strong position dependence of magnitudes is not 
visible in other EFOSC2 data we obtained in 1992 or 1995.
 
In December 1990, we obtained $B,V$ CCD data of NGC 1818 in
direct imaging mode at the same telescope (Will et al.\ 1995a,b). 
The photometry tables are available from CDS, Strasbourg.  The fields 
from the two observing runs largely overlap.  We have used approximately 
2000 common stars from both runs to create correction maps for a 
position-dependent magnitude correction for $BV$ photometry of EFOSC2 
(Grebel \& Will, in prep.).  For NGC 1818, we are using the $BV$ 
photometry of the 1990 run (or $BV$ photometry from 1993 corrected 
accordingly for stars not detected previously because of poorer seeing) 
together with $BVR$ photometry from 1993 and $H\alpha$ photometry from 1992.  
There is an epoch difference of about 20 months between the $H\alpha$ and 
the $R$ frames for NGC 1818. Together with instrumental differences and the 
variability of the Be star phenomenon this may increase the possibility of 
not detecting weak H$\alpha$ emitters and makes the $H\alpha$ calibration 
less accurate.

\begin{table}[thp]
\caption[]{\label{tobslogbe}
Log of observations of NGC 1818 at ESO, La Silla.
For a log of the direct imaging data, see Will et al.\ (1995b).
}
\begin{center}
\scriptsize
\begin{tabular}{cccrcc}
\hline \noalign{\smallskip} 
Instrument     & date        & filter    & exp.time     & airmass & seeing  \\
\noalign{\smallskip}     \hline  \noalign{\smallskip}
2.2m,EFOSC2    & 1993 Oct 08 & $B$         & 40s            & 1.25    &$1\farcs3$\\
               &             & $B$         & 360s           & 1.25    &$1\farcs4$\\
               &             & $V$         & 10s            & 1.26    &$1\farcs1$\\
               &             & $V$         & 120s           & 1.26    &$1\farcs2$\\
               &             & $R$         & 10s            & 1.26    &$1\farcs1$\\
               &             & $R$         & 120s           & 1.26    &$1\farcs2$\\
NTT,EMMI       & 1992 Feb 03 & $H\alpha$ & $3\times 300$s & 1.26    &$1\farcs2$\\
\noalign{\smallskip}     \hline  
\end{tabular}
\normalsize
\end{center}
\end{table}

\section{General properties and colour-magnitude diagram
of NGC 1818}\label{sect_n1818}

NGC 1818 is one of the young, blue, populous clusters in the LMC 
and the central cluster of Region C of the ESO
Key Programme for the Coordinated Investigation of Selected Regions
in the Magellanic Clouds (de Boer et al.\ 1989). 
There have been several spectroscopic abundance determinations for NGC 1818.
Their results differ by more than half a dex in [Fe/H].  The most recent 
studies are both based on medium-resolution spectra of five red supergiants: 
Meliani et al.\ (1994) find NGC 1818 to be relatively metal-poor ([Fe/H] = 
--0.9 dex), while Jasniewicz \& Th\'evenin (1994) find it more metal-rich 
(--0.4 dex).  The latter result may be the more reliable one since care was 
taken to eliminate contamination by blue stars that may lead to spuriously 
low metal abundances. For a summary of earlier spectroscopic abundance 
determinations based on fewer stars see Will et al.\ (1995a).  

\begin{figure}
\centerline{\vbox{
\psfig{figure=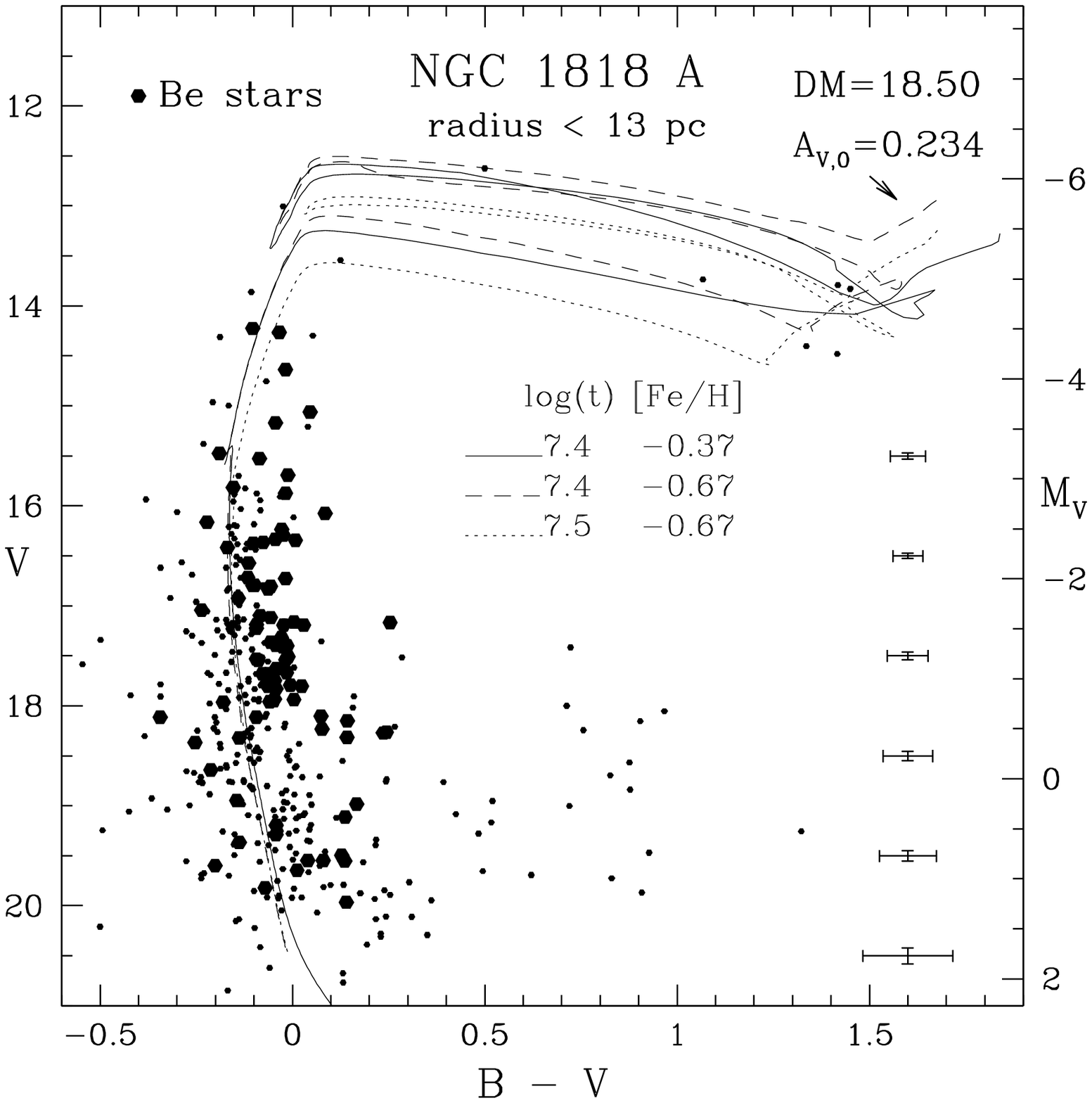,height=6.1cm,width=8.5cm}
\psfig{figure=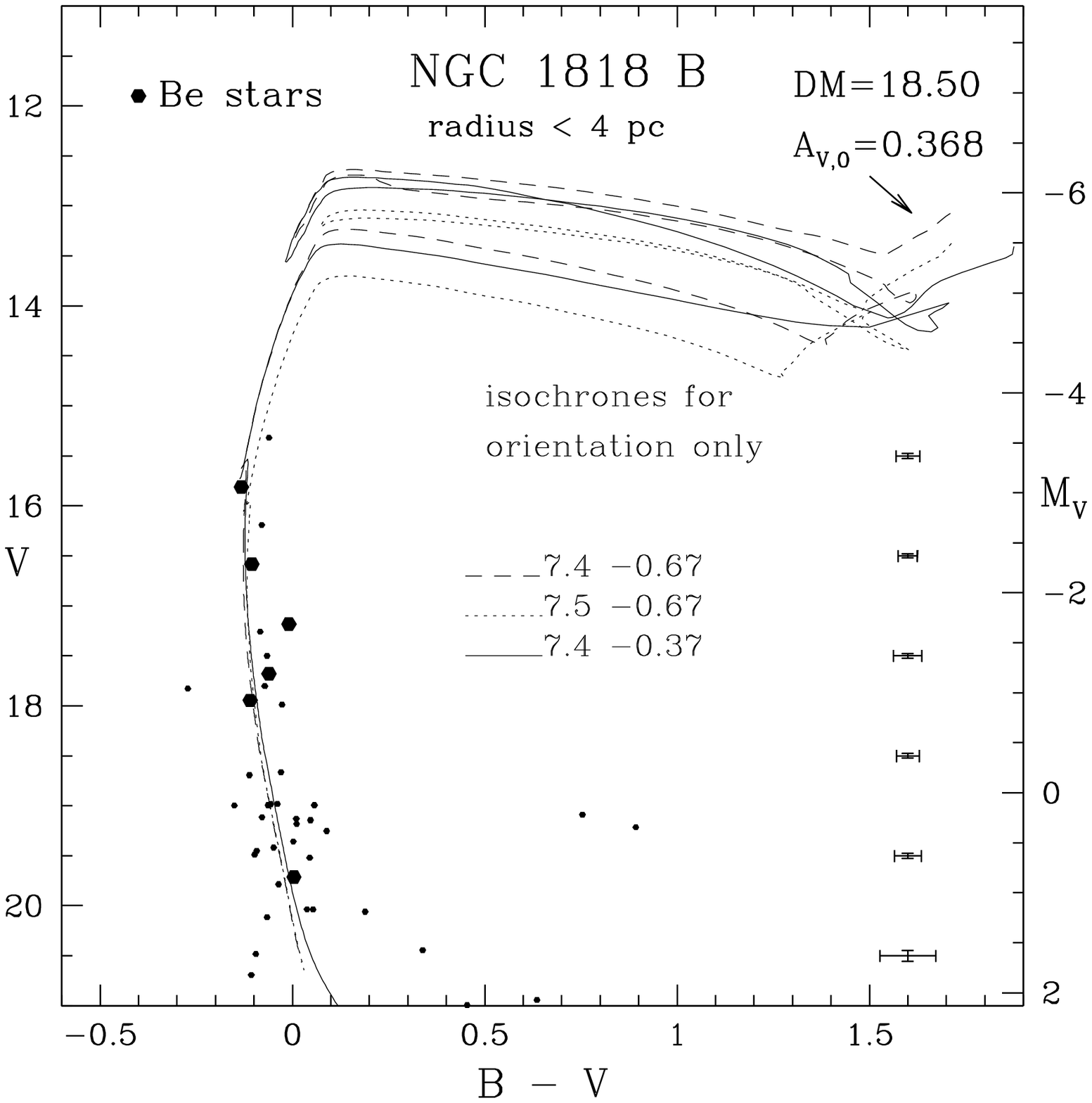,height=6.1cm,width=8.5cm}
\psfig{figure=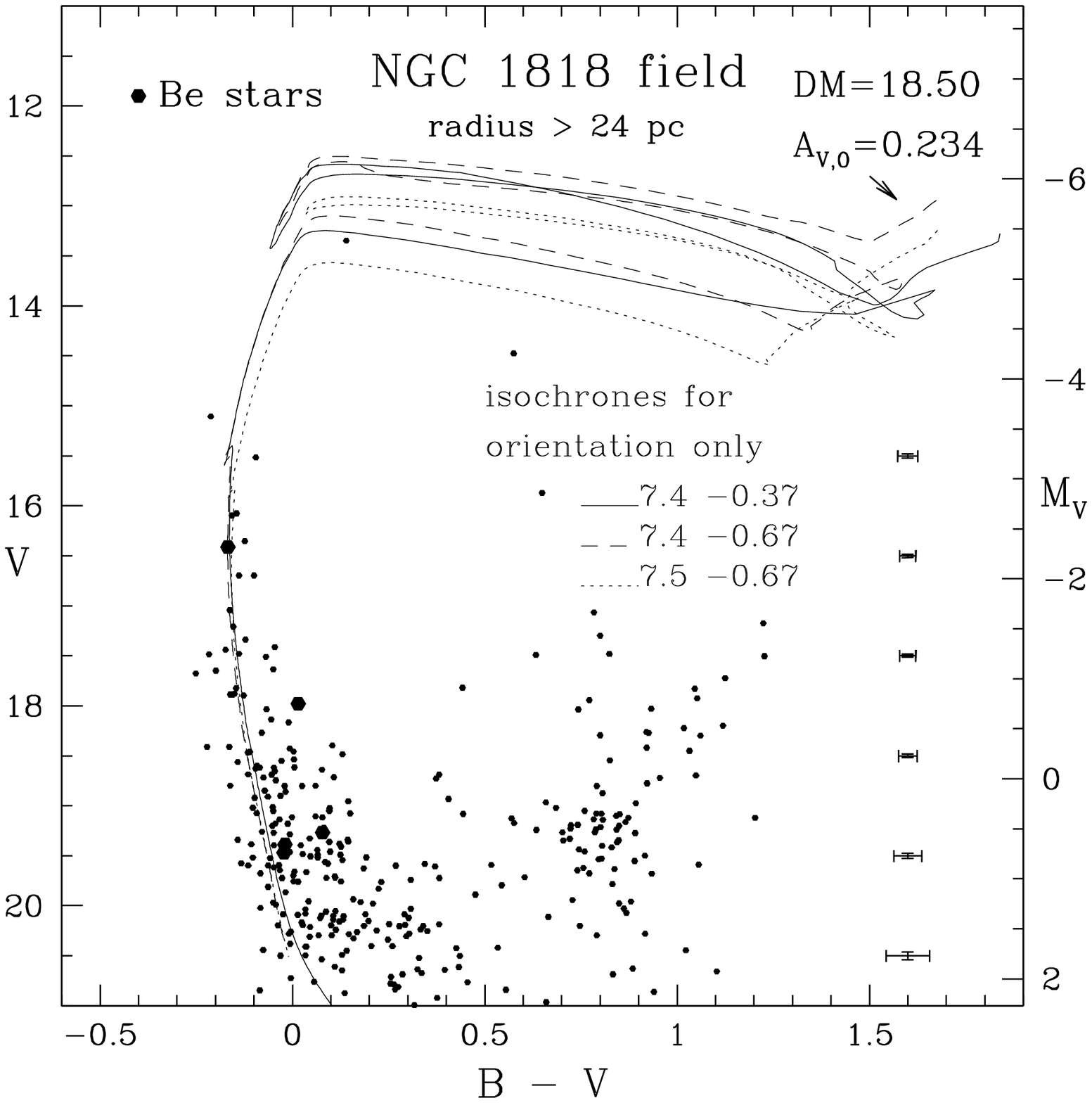,height=6.1cm,width=8.5cm}
}}
\caption[]{ \label{5108F1} 
CMDs of NGC 1818 (upper panel), small cluster (middle), and field population 
(bottom panel). Radii given in the top and bottom panels refer to the centre 
of NGC 1818, and in the central panel to the small cluster. Error bars were 
determined for each region individually.
Geneva group isochrones fit NGC 1818 well for ages between 25 and 30 Myr, 
while showing little sensitivity to metallicity. Isochrones in the central 
and bottom panels are meant for orientation only. 
The small cluster is more heavily reddened than NGC 1818 (see reddening 
vectors)
and does not contain as massive or evolved stars. The field 
main sequence becomes more densely populated at $V\ge18.5$.
The two clusters 
have a high fraction of Be stars (fat dots), but there are
very few in the field. 
}
\end{figure}

The colour-magnitude diagram (CMD) of NGC 1818 shows a very wide blue
main sequence and a few red and blue supergiants (Fig.\ \ref{5108F1}a). 
In Fig.\ \ref{5108F1}a
only stars within 13 pc ($55''$) from the cluster centre are plotted. 
This radius encircles the highest concentration of stars per unit area
before the density distribution becomes noticeably flatter. Thus we may
assume to enclose mostly cluster members.

\begin{figure}
\centerline{\vbox{
\psfig{figure=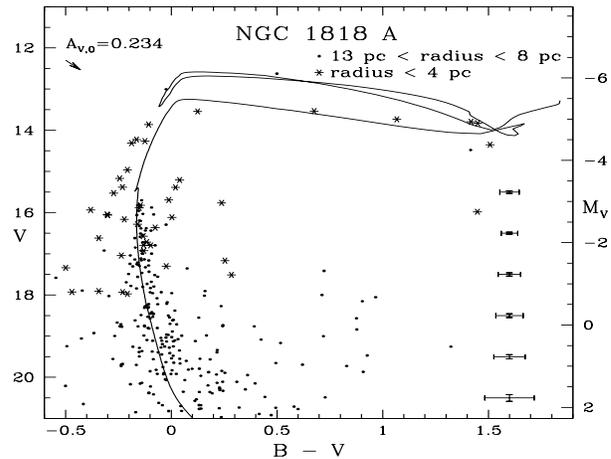,height=6.2cm,width=8.5cm}
}}
\caption[]{ \label{5108F2} 
To illustrate that the very wide main sequence of NGC 1818 is caused
mainly by crowding effects, we compare stars found within an annulus
of 13 pc $<$ radius $<$ 8 pc (small dots) with stars found within
the innermost 4 pc around the cluster centre (asterisks).
The outer annulus of stars shows a quite well-defined, narrow main
sequence. The innermost stars exhibit wide scatter and a much lower
cutoff magnitude.  Scatter toward the blue may be caused in part 
by contamination from unresolved blue main-sequence stars. Note how
strongly the most massive stars are concentrated toward the cluster
centre.
}
\end{figure}

Though the formal photometric colour error bars for NGC 1818 are quite wide
the main sequence still seems excessively wide even when omitting the Be 
stars. Possible reasons for this width are crowding and/or differential 
reddening. Both the error bars and main sequence of the uncrowded field
population are much narrower. In Fig.\ \ref{5108F2}, we plotted stars found 
in an outer and an inner annulus around the centre of NGC 1818. The outermost 
annulus (of stars considered to be largely cluster members, see above) shows 
a relatively narrow, well-defined main sequence for which the isochrones serve 
well as an approximate blue envelope with our chosen reddening. 
In the innermost circle, stars fainter than $V=18$ were not found due to
crowding, and the main sequence exhibits wide scatter. 
The scatter may in part be due to contamination 
by unresolved, fainter blue main-sequence stars. The strange bifurcation
pattern at the upper end of the main sequence 
already visible in the study by Will et al.\ (1995a) originates 
predominantly from these innermost, very crowded stars and would probably 
vanish in high-resolution data such as HST imaging. Fig.\ \ref{5108F2} 
also illustrates that the most massive stars (in particular, supergiants)
are strongly concentrated toward the cluster centre.  

A comparison of stars in different regions within the 13 pc radius
around NGC 1818 did not show significant colour offsets indicating
that effects of differential reddening across NGC 1818 are small if
present at all.

Age determinations based on isochrone fits to colour-magnitude diagrams
were performed by Will et al.\ (1995a, cf.\ also for older references), 
who found ages ranging from 20 Myr to 40 Myr depending on the amount of 
overshoot. From isochrone fits based on models of the Geneva group 
(Charbonnel et al.\ 1993, Schaerer et al.\ 1993) we find an age between 25 
and 30 Myr for NGC 1818 (Fig.\ \ref{5108F1}a). We used a reddening of 
$E(B-V)_0 = 0.07$ mag (Will et al.\ 1995a, Grebel \& Roberts 1995) and a 
distance modulus ($DM$) of 18.5 mag for the LMC (intermediate between recent 
distance determinations, e.g., McCall 1994, Gould 1995). The blue stars 
above the main-sequence turnoff as given by the isochrones and below the 
position of the blue supergiants are in part Be stars, and possibly in part
blue stragglers and/or binaries (for a thorough discussion, see Grebel
et al.\ 1996), and, as just demonstrated, are strongly affected by 
crowding (Fig.\ \ref{5108F2}). 

\subsection{The field population(s)}

We consider stars more distant from NGC 1818 than 24 pc ($100''$)
to be mostly field stars. The field around NGC 1818 contains a number of 
young, massive stars (Fig.\ \ref{5108F1}c and spectral classifications by 
Sanduleak 1970 and Xiradaki et al.\ 1987). These stars are sparsely 
distributed across the field. In particular, there are very few supergiants. 
However, the main sequence becomes more densely populated starting at $V 
\approx 18.5$. From B0 V to B5 V, $M_V$ decreases by approximately 2.8 mag, 
while $M_{\rm bol}$ decreases by roughly 1.1. Assuming an IMF power law
as given by, e.g., Scalo (1986) with a Salpeter slope of 1.35 (Salpeter 1955), 
the number of stars within the above brightness interval should 
increase by about 1.5, or by about 1.4 when using the IMF slope
determined for NGC 1818 by Will et al.\ (1995a). The number of stars
increases by more than that though.
The onset of higher stellar densities corresponds to an intermediate-age
population with an age of 400 Myr and older. The lower main sequence
becomes quite wide, which may be due to differential reddening and/or
depth effects. The blue envelope of the main sequence fits isochrones
with a reddening of $E(B-V)=0.07$ mag.  
The field CMD also shows the typical clump of (older) intermediate-age
red giants at $V\approx 19$. Older populations cannot be discerned in our
data though since we do not reach faint enough magnitudes.

In the field surrounding NGC 1818 there are four more clusters located, 
all less populous than NGC 1818. These are SL 205 (northeast at a distance
of $5\farcm4$ or 78 pc), NGC 1810 (northwest at a distance of $6\farcm2$
or 90 pc), an anonymous cluster (southwest at a distance of $6\farcm6$
or 95 pc), and SL 222 (southeast at a distance of $8\farcm3$). The age
of NGC 1810 was estimated to be 76 Myr, and its reddening to be higher 
($E(B-V) = 0.12$ mag) than that of NGC 1818 (e.g., Meurer et al.\ 1990).
No data on the other three clusters are available in the literature.
Judging from the corresponding plate scan available in the STScI Digitized Sky
Survey, SL 205 appears to be associated with a gaseous region and may
be rather young. Judging from these clusters and the scattered supergiants
in the field, widespread though sparse star formation must have taken place 
across the entire region within the past 25 to 80 Myr.  The variations in 
reddening make the presence of differential reddening seem likely.

\subsection{A companion cluster of NGC 1818?}\label{sect_n1818B}

About $1\farcm5$ southwest of NGC 1818 there is a second, very small
cluster (see Fig.\ 1 in Will et al.\ 1995b). It is uncertain whether 
this cluster is a binary companion of NGC 1818 or if it is seen in
chance projection.  For convenience, we will refer to 
the main cluster as NGC 1818 or NGC 1818A and to the small cluster as NGC 
1818B. 

The CMD of NGC 1818B (Fig.\ \ref{5108F1}b) shows a sparsely populated main 
sequence that does not extend to stars as massive as found in NGC 1818A. 
There are no evolved (i.e., supergiant) stars present in the small cluster. 
The lack of more massive stars may be due to small number statistics, 
a less massive birth cloud or subcloud, or due to disruption of star 
formation (Franco et al.\ 1994) caused by NGC 1818A if the clusters are 
associated.  The apparent absence of evolved stars and the very sparsely
populated main sequence make it difficult to assign an age to NGC 1818B. 
Judging from the two brightest main-sequence stars, it {\em may} be coeval 
with NGC 1818A. 

Plotting isochrones onto the CMD of NGC 1818B indicates that the reddening 
of this cluster is higher than that of NGC 1818A, namely $E(B-V) \approx 0.11$ 
mag. This may be due to the presence of differential reddening in this area, 
or a larger distance and thus higher reddening for NGC 1818B. It is also 
possible that star formation was less efficient for NGC 1818B and that there 
is still a higher concentration of dust and gas in this area. The currently 
available data do not yet allow to conclude whether the small cluster is 
associated with NGC 1818A. Radial velocities and spectral types for the 
brightest stars would help to clarify this matter.

\section{Be stars in NGC 1818 and neighbouring cluster}\label{sect_n1818be}

Detections of Be stars in NGC 1818 were simultaneously reported by
Bessell \& Wood (1993) and Grebel et al.\ (1993). Both studies use the
photometric detection method described in Sect.\ \ref{sect_intro}.
An early attempt to estimate and compare Be star fractions was performed 
by Grebel et al.\ (1994). The current study gives a table with
photometry and coordinates of the detected Be stars as well as a detailed
comparison with Be stars in other Magellanic Cloud clusters and Galactic
open clusters.  

\begin{figure}
\centerline{\vbox{
\psfig{figure=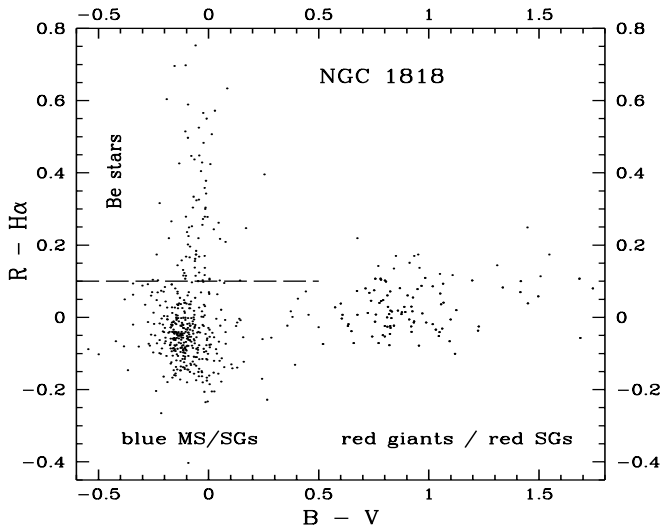,height=6.8cm,width=8.5cm}
}}
\caption[]{ \label{5108F3} 
Two-colour diagram for the detection of Be stars. As selection criterion
for Be stars we use ($B-V$)$\le 0.5$ mag and ($R-H\alpha$)$\ge0.1$ mag (dashed
line).
}
\end{figure}

To detect H$\alpha$-emitting stars, we plot ($R-H\alpha$) versus ($B-V$)
(Fig.\ \ref{5108F3}). Stars without H$\alpha$ emission can be found around 
($R-H\alpha$)$\approx0$. Stars with H$\alpha$ emission have ($R-H\alpha$) 
indices larger than zero.  A pronounced clump of stars at ($B-V$) $\le-0.2$ 
mag and ($R-H\alpha$) $\loa0$ mag can be seen. These stars are blue 
main-sequence stars and blue supergiants that do not currently show H$\alpha$ 
emission.  Redward of this clump there is an extended group of stars scattered 
around ($R-H\alpha$)$\approx0$ mag. These stars are red giants belonging to 
the field population and red supergiants from NGC 1818A. Red giants or 
supergiants may show H$\alpha$ emission, and there are indeed a few red stars 
bright in H$\alpha$.  The third group of stars lies along a band stretching 
upwards to positive values of ($R-H\alpha$) starting from the clump of blue 
main sequence stars and supergiants. These stars are blue stars with H$\alpha$ 
emission, i.e., our Be star candidates. When a larger colour baseline
involving red filters is plotted such as ($B-R$) or ($V-I$) is used, one can 
see that this band of stars is tilted such that the stars brightest in 
H$\alpha$ tend to be redder than the fainter emission-line stars (see figures 
in Grebel et al.\ 1994). 

Due to the scattered clump of main-sequence stars and supergiants 
around ($R-H\alpha$)=0 mag we adopt ($R-H\alpha$)$>0.1$ mag 
as selection criterion for Be star candidates. A less strict
criterion would increase the number of possible weak-line Be
star candidates but would also increase the probability of including
non-emission line stars.  As mentioned before, reliable detections
of very faint H$\alpha$ emitters are only possible spectroscopically
(see also Sect.\ \ref{sectbe_prob}).

The Be star candidates that are the most luminous at H$\alpha$ can be 
found at ($R-H\alpha$)$\approx 0.7$ mag. In Tab.\ 2 (available from CDS)
we list 
$BVRH\alpha$ photometry of Be stars in NGC 1818 (marked by an ``A'') and 
the small nearby cluster (marked by a ``B'').  The field (marked by ``F''s) 
around NGC 1818 and contains almost no Be stars. The $V$ luminosity of the 
two or three brightest Be stars in our list indicates that they may qualify 
as extreme Be stars (see Garmany \& Humphreys 1985 for more information). 
Spectroscopic follow-up studies are needed to confirm or reject this
possibility. The astrometric data, right ascensions and declinations for 
equinox J2000.0, were taken from Will et al.\ (1995b) or, where not available, 
interpolated using the stars from Will et al.\ as tertiary astrometric grid. 
The astrometric positions from Will et al.\ (1995b) themselves are based on 
the grid of secondary astrometric reference stars in the Magellanic Cloud 
Catalogue of Stars (MACS, Tucholke et al.\ 1996) and have an accuracy of 
$0\farcs5$ and $0\farcs6$.

Our identifiers for the Be stars comply with the specifications for
designations of astronomical objects as suggested by IAU Commission 5
(http://astro.u-strasbg.fr/iau-spec.html). The identifier
``NGC 1818:GBe xxx'' specifies the source (NGC 1818) within and around which
the Be stars are located. Separated through a colon are the subcomponents,
which contain the letter sequence (GBe) and three digits (xxx) to identify
the individual star. ``GBe'' consists of the author's initial (G) and the
string ``Be'' to indicate the nature of the stars.

\section{Be star fractions}\label{sectbe_prob}
									 
In Grebel et al.\ (1994) we investigated the fraction	
of Be stars in comparison to ordinary B stars using simple	
magnitude bins whose width in $V$ magnitude corresponded to spectral subtypes
as tabulated by Schmidt-Kaler (1982). The Galactic Be star frequencies	
were taken from Jaschek \& Jaschek (1983).
However, this approach is not entirely correct.
We compared Be stars of supposedly common origin and supposedly	
similar age and chemical composition (namely Be stars in a specific
cluster) to Galactic field Be stars, thus Be stars of quite different
age, composition, and formation conditions. For a valid comparison,  
Be star fractions in Milky Way and in Magellanic Cloud clusters should be 
compared. Lacking blue populous clusters in the Milky Way, young open 
clusters are best suited. 

Secondly, the distribution of Be stars with spectral type and their overall
number depends on the age of the parent cluster (Mermilliod 1982)	
showed. The highest Be star frequencies were found in clusters with turnoffs
in the spectral region of B0.5 to B2 stars. This corresponds to the 
ages of the three Magellanic clusters that we have analyzed so far -- 	 
NGC 330, NGC 2004, and NGC 1818 (Grebel et al.\ 1992, 1994, 
Grebel 1995, and this study).  
Thus, for a meaningful comparison our data must be compared to Galactic	 
clusters of the same age group.	

The detection of Be stars is, of course, affected by incompleteness.
For a given cluster we are less likely to detect Be stars of later spectral 
type because those stars will generally have fainter magnitudes.
Generally, Balmer emission is strongest for the earliest types of Be
stars. Also, in a crowded region detections are biased toward the
brighter stars (affecting, of course, both B and Be stars).
As illustrated in Fig.\ \ref{5108F2}, we are severely affected by
incompleteness in the innermost areas of NGC 1818. Close to
the detection limit Be stars are more likely to be detected in H$\alpha$
than non-emission line B stars since they may be up to several
tenths of a magnitude brighter. Stars with very faint Balmer emission
can only be detected spectroscopically anyway (see Sect.\ \ref{sect_intro}).
 
The fact that B-type stars may appear as Be stars at certain times and as 
ordinary B stars at others affects the Be star census in Galactic open 
clusters and star clusters in the Magellanic Clouds equally.  It is important 
to remember that all numbers represent only lower limits.

The Galactic open clusters with the largest known Be star fractions are NGC 
663, NGC 3766, and $\chi$ Persei (e.g., Slettebak 1985). The first two 
clusters also have similar ages (a few $10^7$ years) as the Magellanic Cloud 
clusters.  The total number of known Be stars in these clusters is 24 for 
NGC 663 (Sanduleak 1990), 30 for h and $\chi$ Per (Slettebak 1985), and 19 
for NGC 3766 (Shobbrook 1985, 1987).  Many of these clusters have been 
observed repeatedly over many years, and spectroscopy is available for most 
of their bright B-type stars. Thus the data on their 
Be stars may be more complete than our Magellanic cluster data,
two of which we observed at only one epoch.
 
Among early-type B stars (B0 to B5), the estimated Be star fractions
are 40\% in NGC 663 (Sanduleak 1990) and 25\% in $\chi$ Per. Based on 
photometric monitoring, Waelkens et al.\ (1990) suggest that up to  
50\% of the B stars in h and $\chi$ Per may be Be stars.  Shobbrook's (1985, 
1987) results indicate a Be star fraction of at least 36 \% for NGC 3766. 
Similarly, Hillenbrand et al.\ (1993) propose that 34 \% of the early-type 
stars they studied may be Be stars.

For the LMC cluster NGC 2004, Kjeldsen \& Baade (1994) carried out slitless
H$\alpha$ spectroscopy and found 43 Be stars in a field of $7' \times 7'$.
They estimate the Be star fraction to be 30\% with a peak at
45\% around B3 to B4, but point out that limited 
photometry and spectral classification in the literature made it difficult 
for them to estimate Be star fractions.  We find an overall larger number 
of Be stars within a smaller field of view. 

\subsection{Assigning spectral types through photometry}\label{secthistphot}

\begin{figure*}
\centerline{\hbox{
\psfig{figure=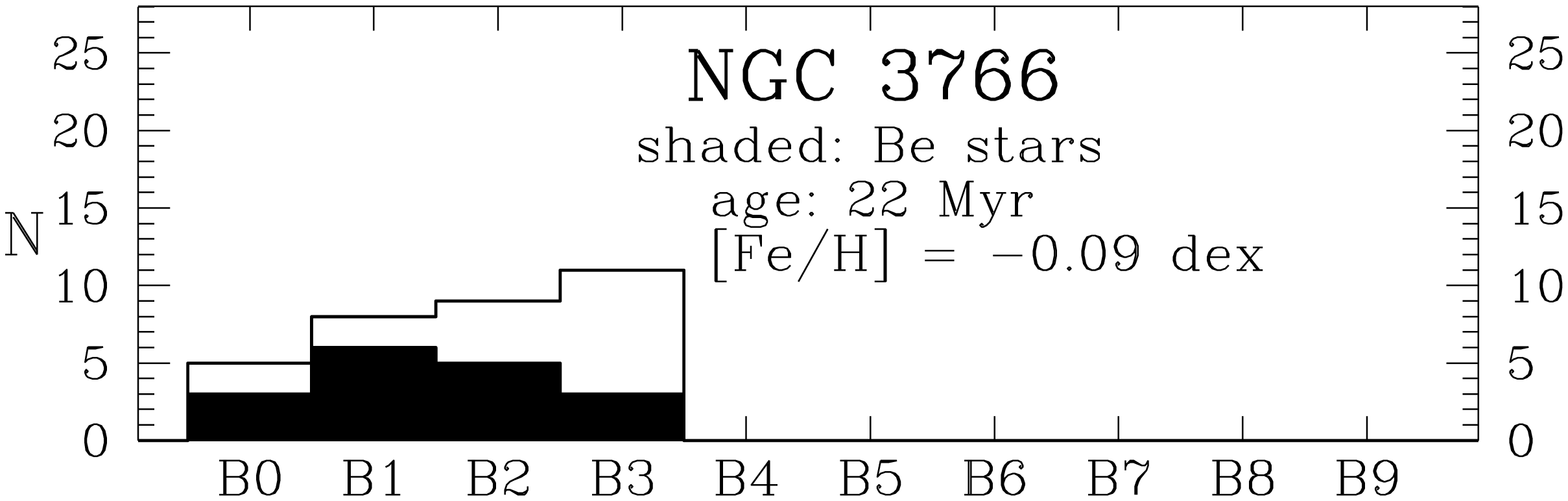,height=2.0cm,width=8.3cm,angle=-90}
\psfig{figure=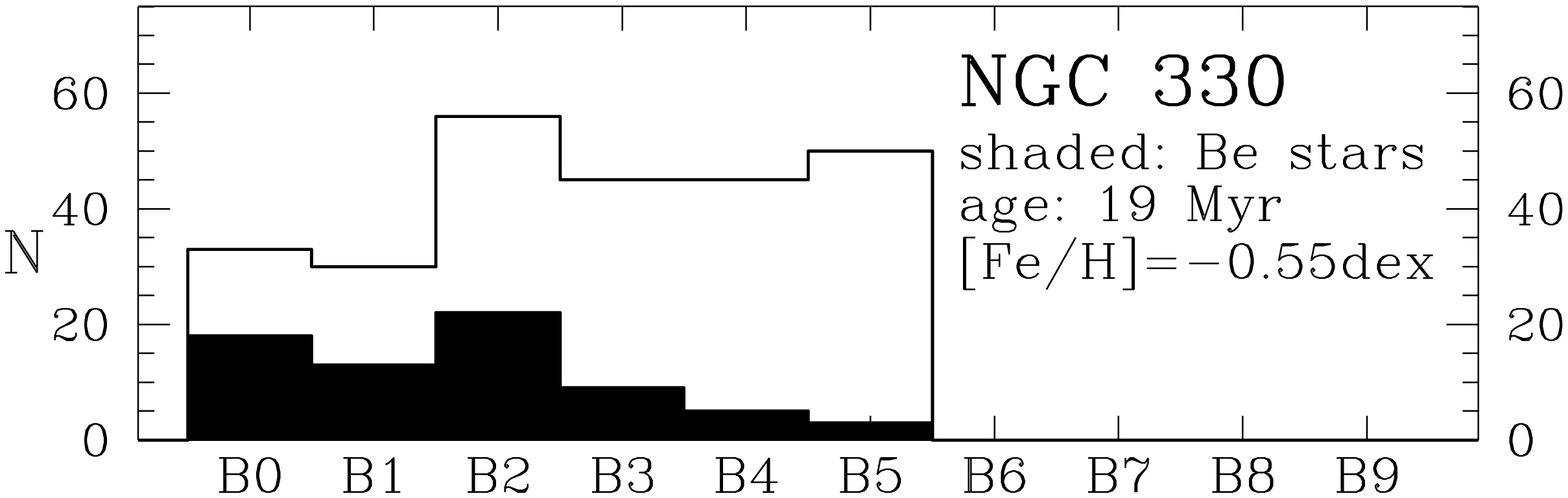,height=2.0cm,width=8.3cm,angle=-90}
}}
\centerline{\hbox{
\psfig{figure=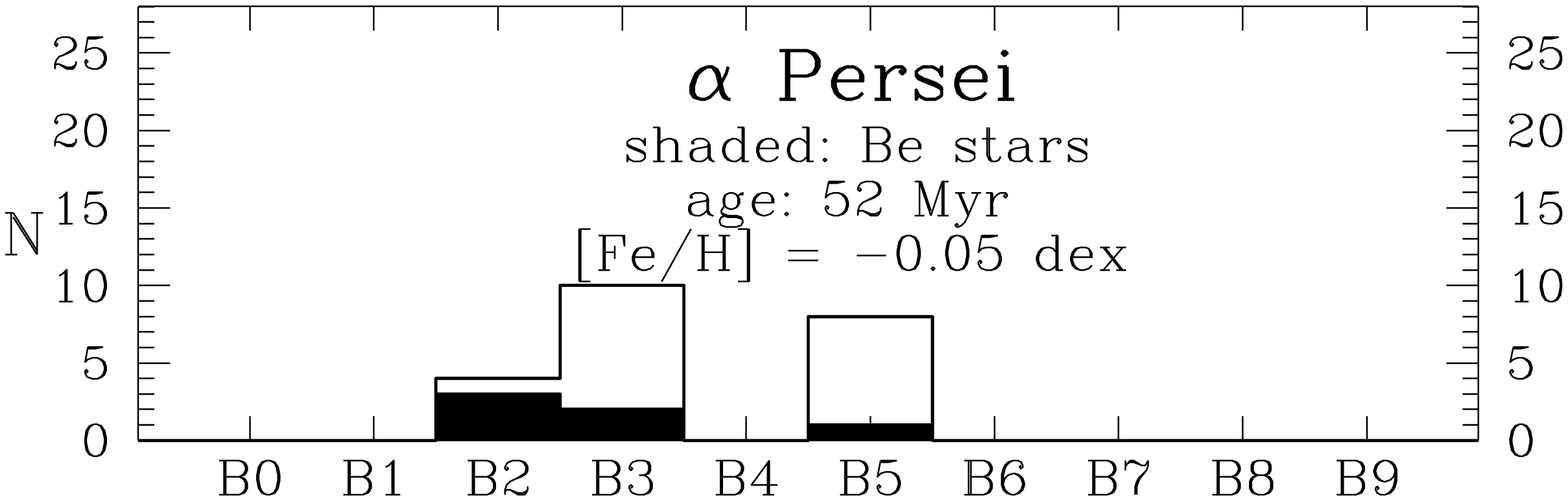,height=2.0cm,width=8.3cm,angle=-90}
\psfig{figure=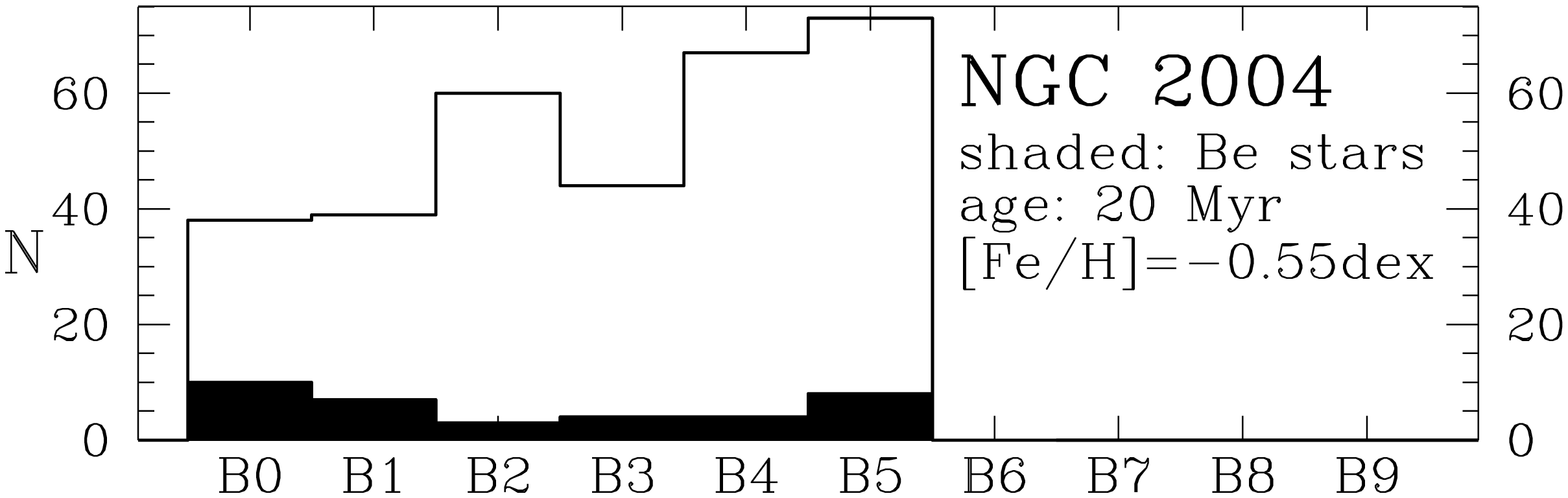,height=2.0cm,width=8.3cm,angle=-90}
}}
\centerline{\hbox{
\psfig{figure=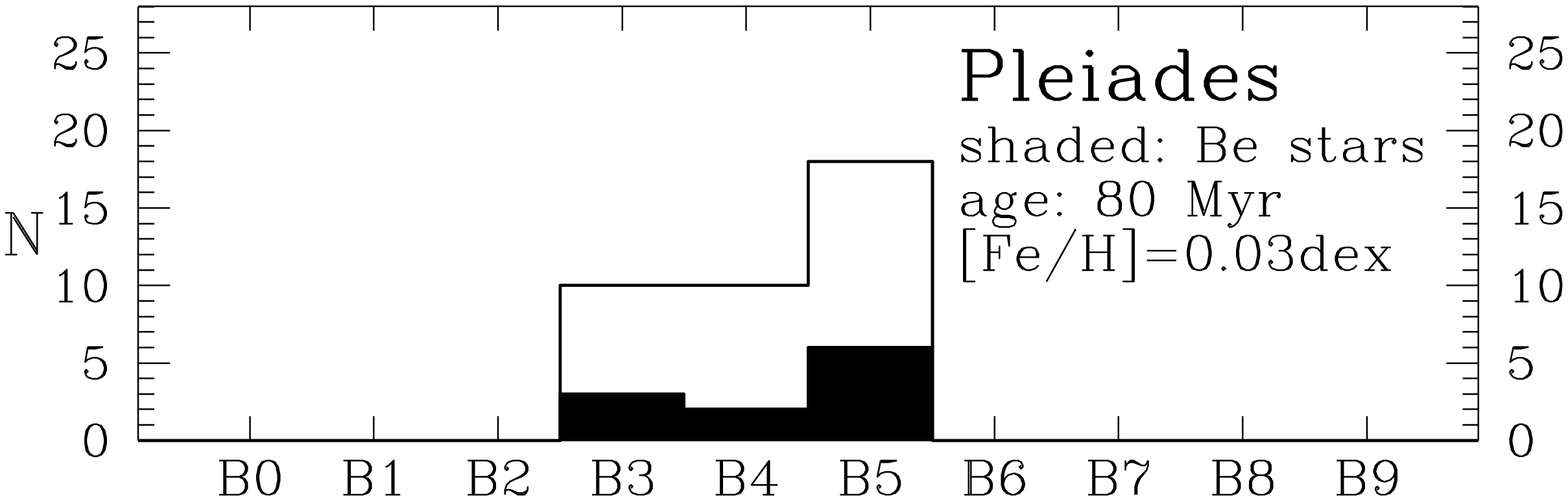,height=2.0cm,width=8.3cm,angle=-90}
\psfig{figure=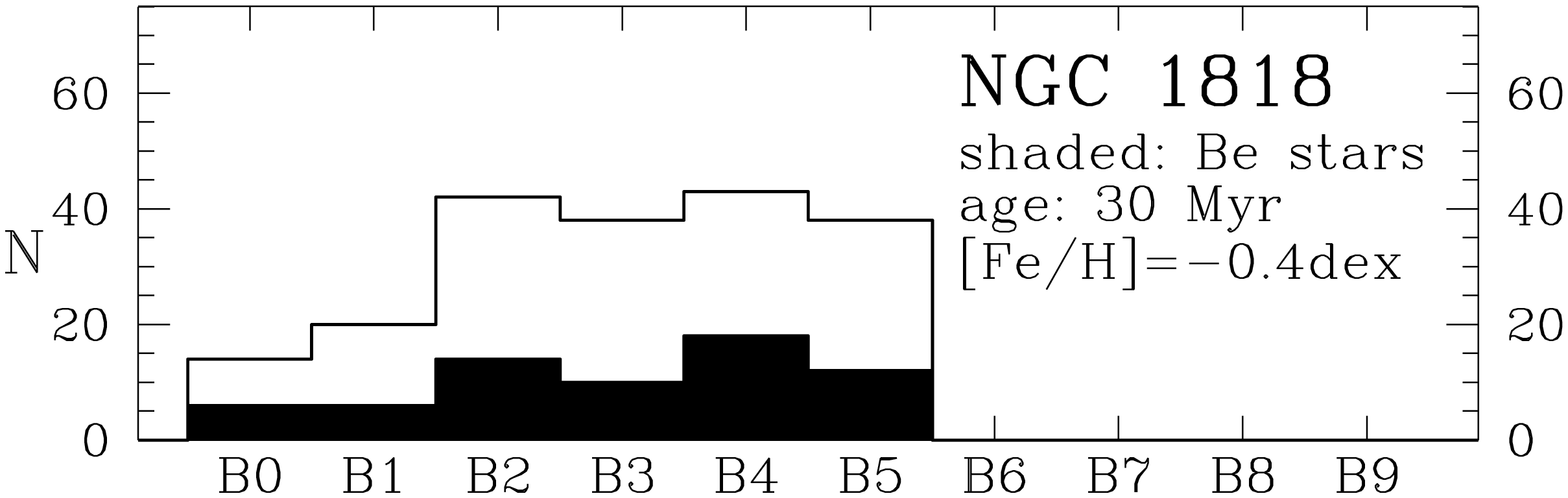,height=2.0cm,width=8.3cm,angle=-90}
}}
\caption[]{ \label{5108F4} 
Spectral types assigned through a photometric calibration derived from
Zorec \& Briot (1991) are plotted for three Galactic open clusters rich in
Be stars (left panels) and our three Magellanic Cloud clusters (right panels).
``N'' denotes total numbers of stars. B stars in general are represented by
white histograms, while the black histograms indicate the number of Be stars
among them. A cutoff toward later B types is artificially introduced through
lack of data. The diagrams seem to indicate that Be star fractions are 
particularly high for the earliest B stars (B0 to B2). However, note the 
significant shifts in spectral types when spectroscopic results are used 
instead (Fig.\ \protect\ref{5108F5}). 
For the Pleiades, spectral typing through photometry moved the luminous
non-supergiants to too early types by up to three B sub-classes. }
\end{figure*}

Since we are lacking adequate spectroscopic data for NGC 1818 and NGC 2004, we 
have tried to assign spectral types to the Be stars through a relationship 
between	absolute visual magnitudes and spectral types. Commonly used 
calibrations of spectral type versus absolute magnitudes such as the 
tabulations of Schmidt-Kaler (1982) do not consider the considerable 
intrinsic dispersion of absolute magnitudes that exists in each magnitude
(Zorec \& Briot 1991). This dispersion has been found to be about one 
magnitude for B2 V stars, and even two magnitudes for B9 III stars 
(Chalonge \& Divan 1973). Zorec \& Briot (1991) performed a new calibration of	
absolute magnitudes and spectral classes for B-type stars and also determined
the scatter in magnitude. They find that Be stars are as luminous or up	
to about 0.8 mag (in $V$) more luminous  than B stars of the same spectral type.
Due to their generally rapid rotation, departures from sphericity may lead
to changes in spectral type, brightening, and intrinsic ``reddening'' effects
including limb darkening and gravity darkening (Slettebak et al.\ 1980,  
Collins et al.\ 1991, see also discussion in Grebel et al.\ 1996). 
As pointed out by Collins \& Smith (1985), the position of an individual 
rotationally displaced star in a CMD cannot be uniquely associated with a 
single position on the zero-rotation main sequence.
Briot \& Zorec (1994) corrected for effects on Galactic field
Be stars introduced by the over-luminosity of Be stars, spectral type
changes during constant mass evolution, and spectral type changes due to
fast rotation. Furthermore, Be stars may be affected by intrinsic reddening 
caused by their circumstellar disks. This effect will again depend on the 
inclination angle. The Be star phenomenon is known to occur as well in binary 
systems. Be stars that are members of binaries will also be displaced from 
the main sequence (see, e.g., Pols et al.\ 1991).

Adopting the spectral class vs.\ $M_V$ calibration {\sl for main 
sequence stars} given by Zorec \& Briot (1991), we approximated the spectral
subtypes for the Be stars as follows: 
$M_{V, \rm B0} < -3.25$, $-3.25 <M_{V, \rm B1} <
-2.55$, $-2.55< M_{V, \rm B2} < -1.8$, $-1.8 <  M_{V, \rm B3} <
-1.4$, $-1.4 < M_{V, \rm B4} < -0.95$, $-0.95 < M_{V, \rm B5} < -0.6$.
Due to increasing photometric errors and field contamination for the 
Magellanic Cloud clusters we do not consider stars of presumably later
types than B5.

Using these magnitude bins, we assigned spectral types to our photometric
data on NGC 1818, NGC 330, and NGC 2004, and to published data on the 
Galactic open clusters NGC 3766, $\alpha$ Per, and the Pleiades. The numbers 
for Galactic open clusters were extracted from dereddened CMDs published by 
Mermilliod (1982) and Slettebak (1985).  Furthermore, we used tabulations by 
Shobbrook (1985, 1987: NGC 3766).  We adopted the distance moduli and 
reddenings given in these references.  The ages of open clusters are from 
Lyng\aa\ (1987: NGC 3766) and Meynet et al.\ (1993: $\alpha$ Per, Pleiades). 
Spectroscopic abundance determinations were taken from Luck \& Bond (1989: 
NGC 3766) and Boesgaard \& Friel (1990: $\alpha$ Per, Pleiades).

For our own data, we used $DM = 18.9$ mag for NGC 330, and $DM = 18.5$ mag 
for NGC 1818 and 2004, and ages and reddenings resulting from best-fitting 
isochrones (Grebel et al.\ 1994, Grebel 1995, and Sect.\ \ref{sect_n1818}).  
The small cluster southwest of NGC 1818 is not included in the comparison 
because of poor number statistics. 

In Fig.\ \ref{5108F4}, we plot our results for the three Galactic open
clusters and the three Magellanic Cloud clusters. 
Comparing the diagrams, the overall fraction of Be stars among the B-type
stars appears to be higher for the three Galactic clusters than for the
the Magellanic clusters. 
The diagrams seem to indicate that Be star fractions are particularly 
high for the earliest B stars, i.e., B0 to B2. As we will show in Sect.\
\ref{secthistspec}, however, assigning spectral types through the photometric
calibration leads to spurious results despite the careful work by Zorec \&
Briot (1991). 

\subsection{Be star fractions from spectroscopy}\label{secthistspec}

\begin{figure*}
\centerline{\hbox{
\psfig{figure=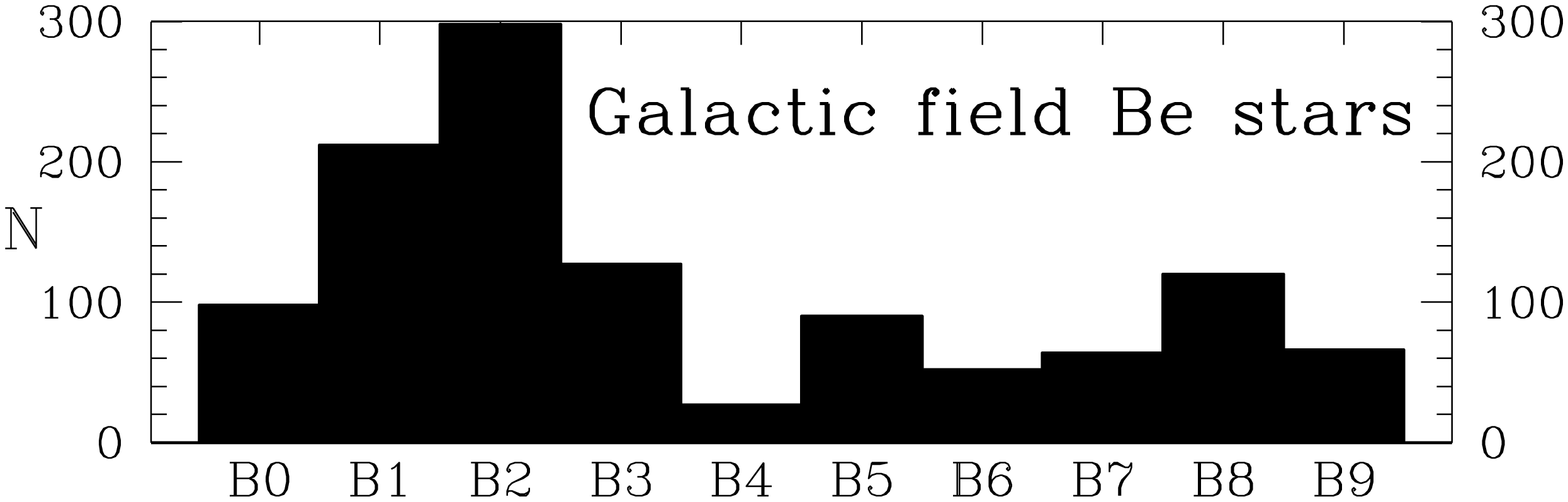,height=2.0cm,width=8.3cm,angle=-90}
\psfig{figure=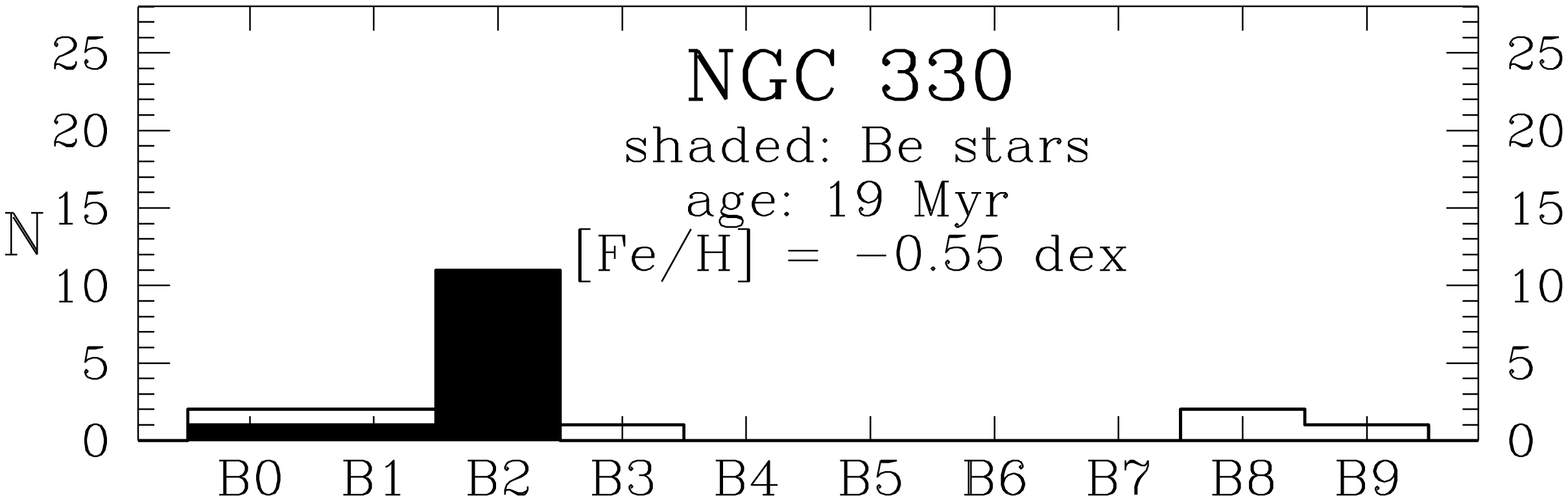,height=2.0cm,width=8.3cm,angle=-90}
}}
\centerline{\hbox{
\psfig{figure=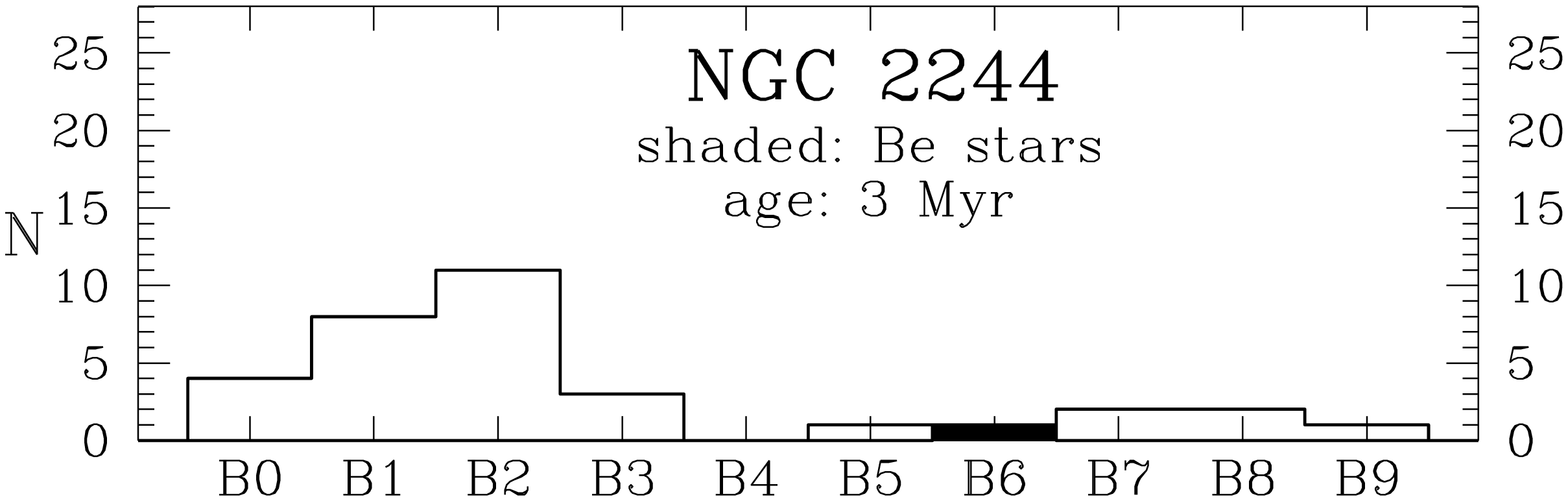,height=2.0cm,width=8.3cm,angle=-90}
\psfig{figure=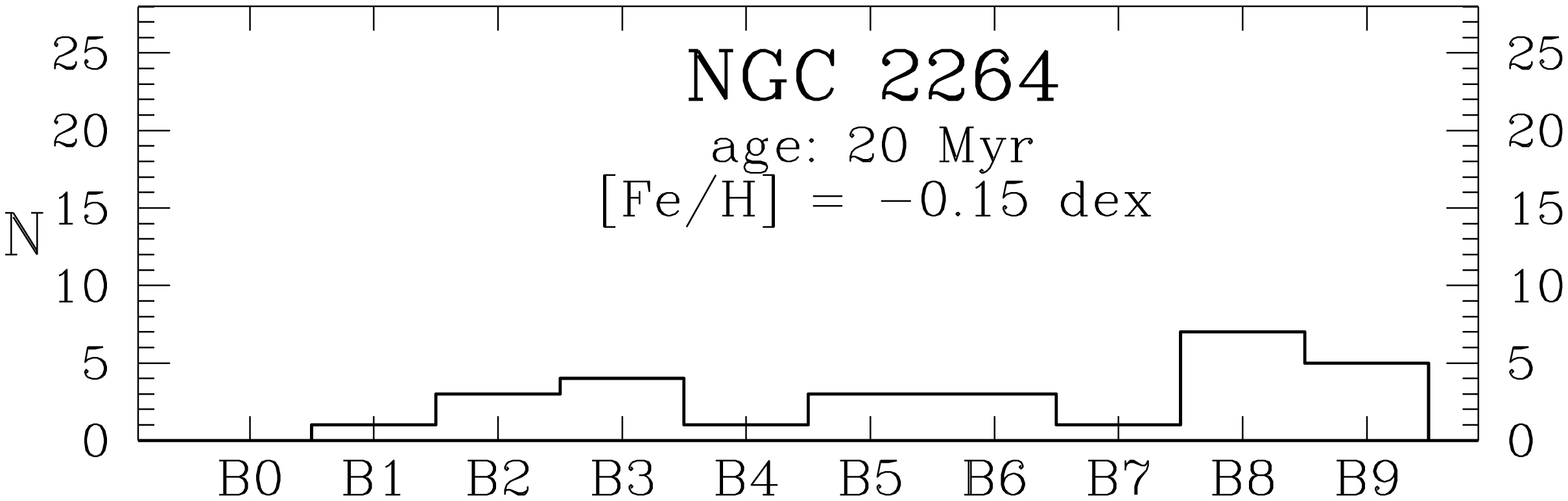,height=2.0cm,width=8.3cm,angle=-90}
}}
\centerline{\hbox{
\psfig{figure=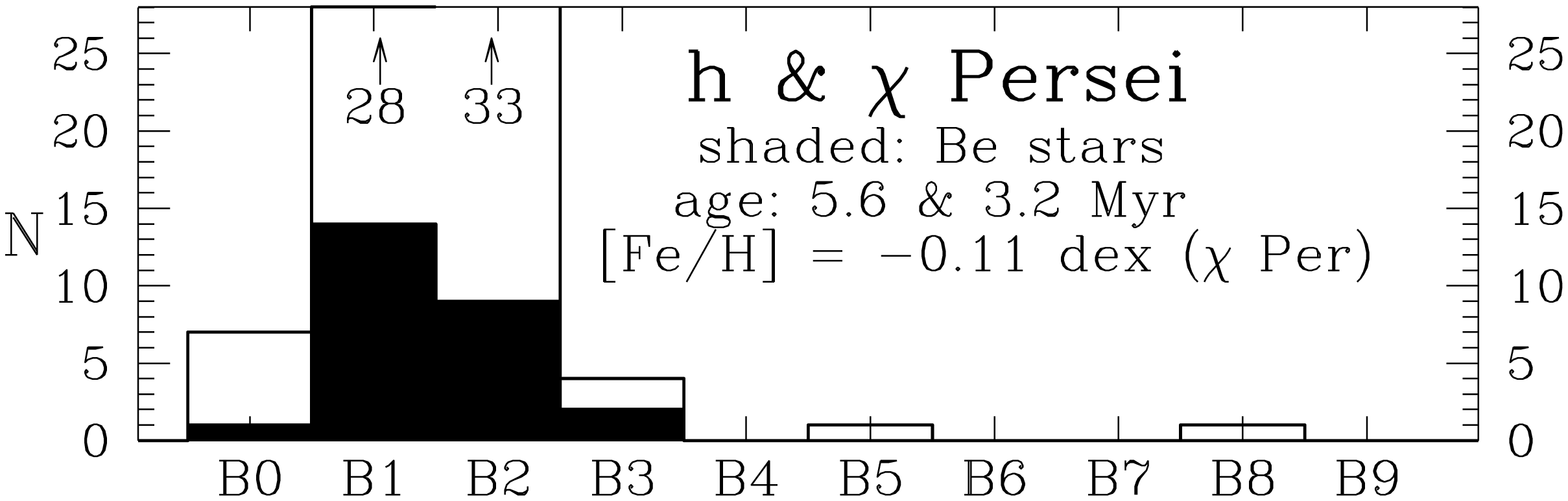,height=2.0cm,width=8.3cm,angle=-90}
\psfig{figure=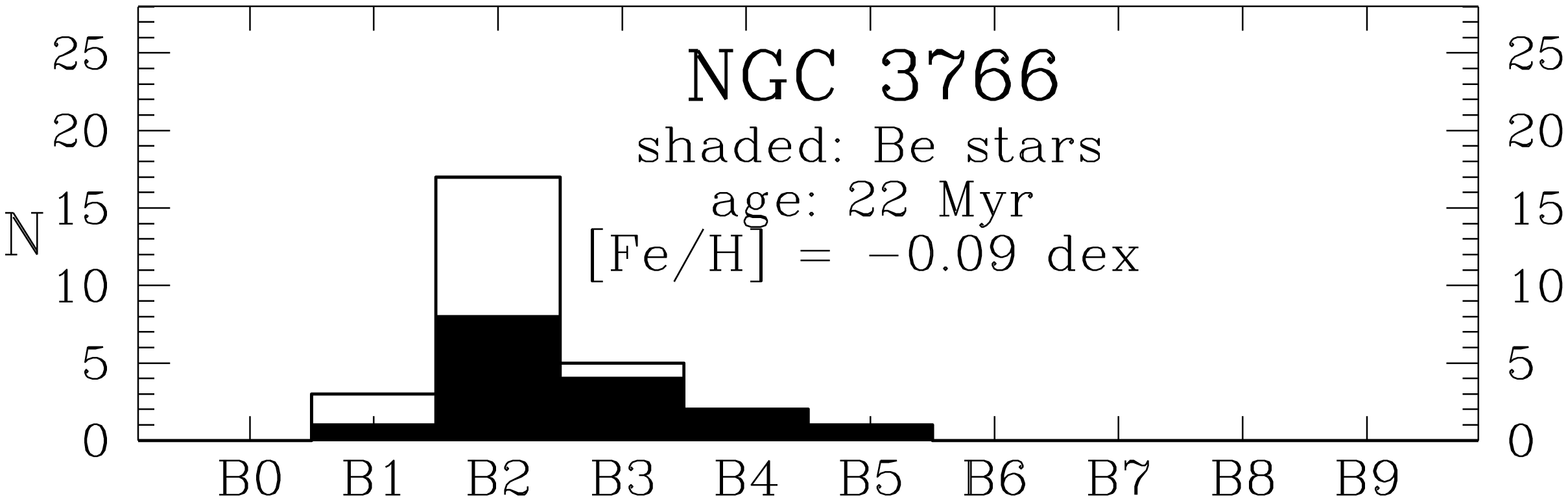,height=2.0cm,width=8.3cm,angle=-90}
}}
\centerline{\hbox{
\psfig{figure=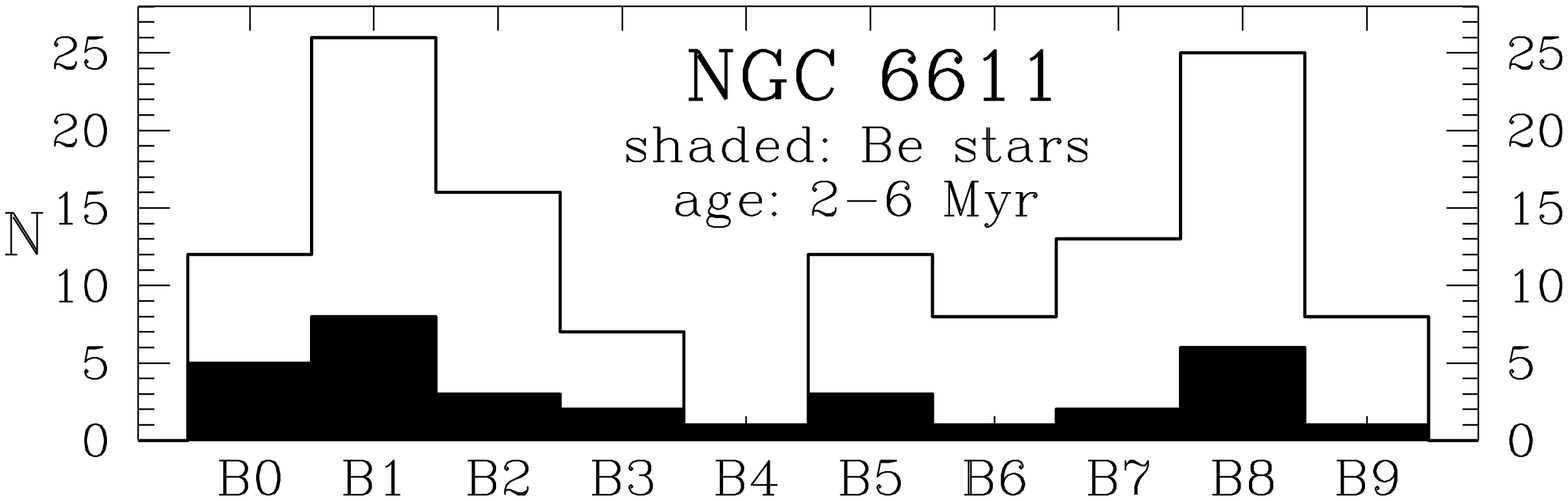,height=2.0cm,width=8.3cm,angle=-90}
\psfig{figure=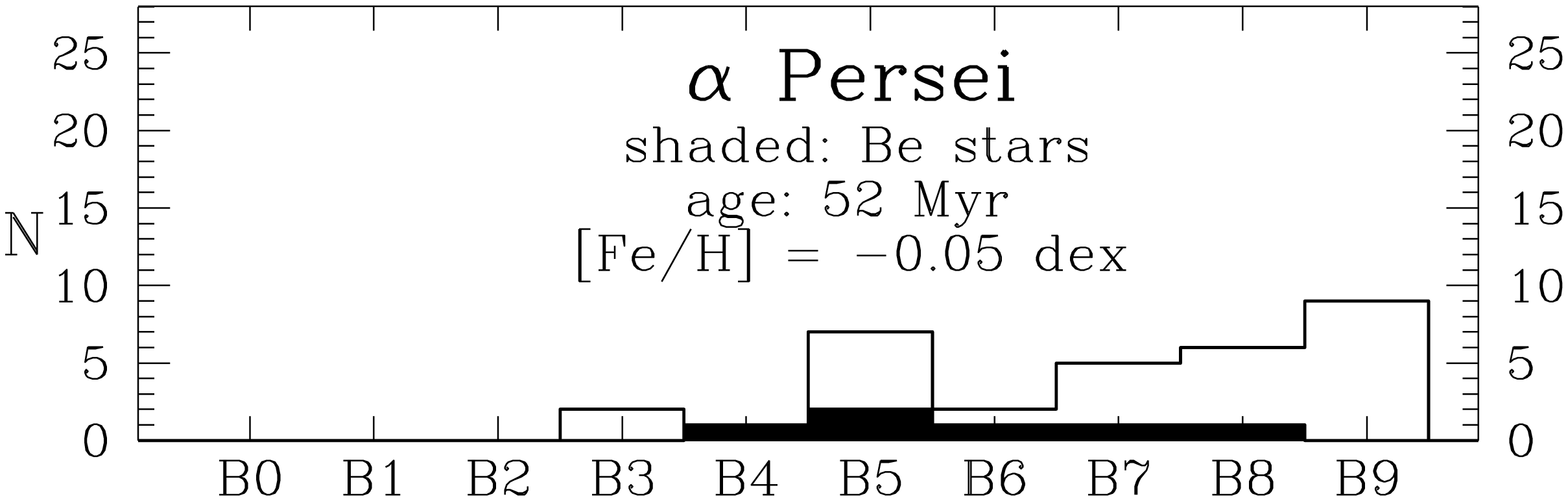,height=2.0cm,width=8.3cm,angle=-90}
}}
\centerline{\hbox{
\psfig{figure=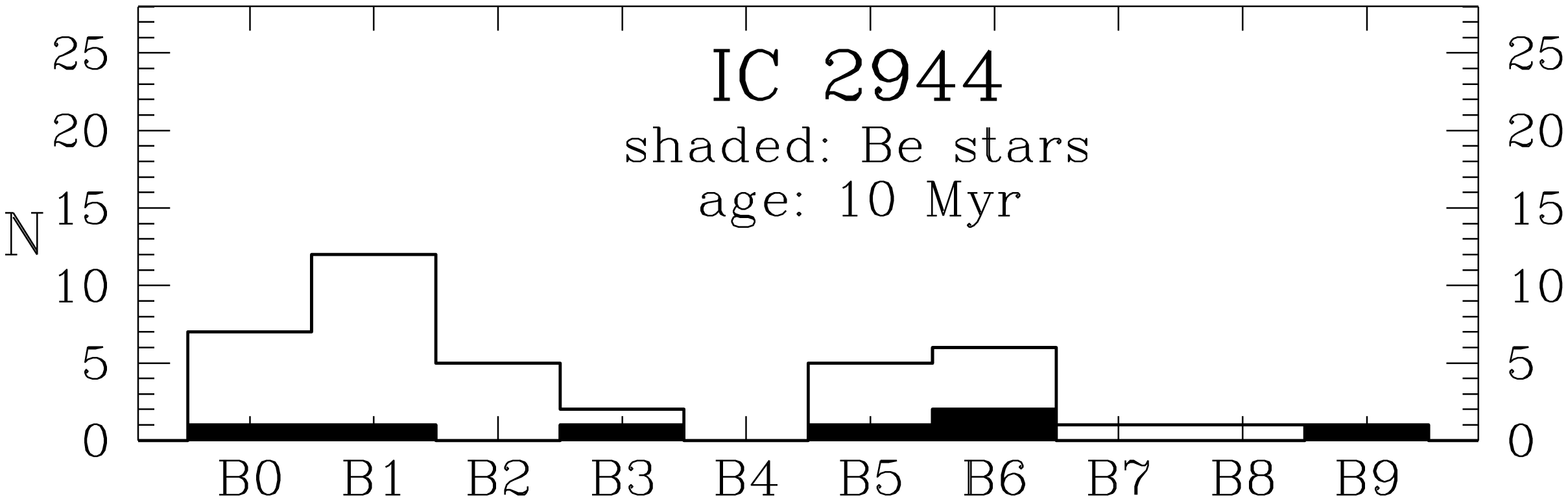,height=2.0cm,width=8.3cm,angle=-90}
\psfig{figure=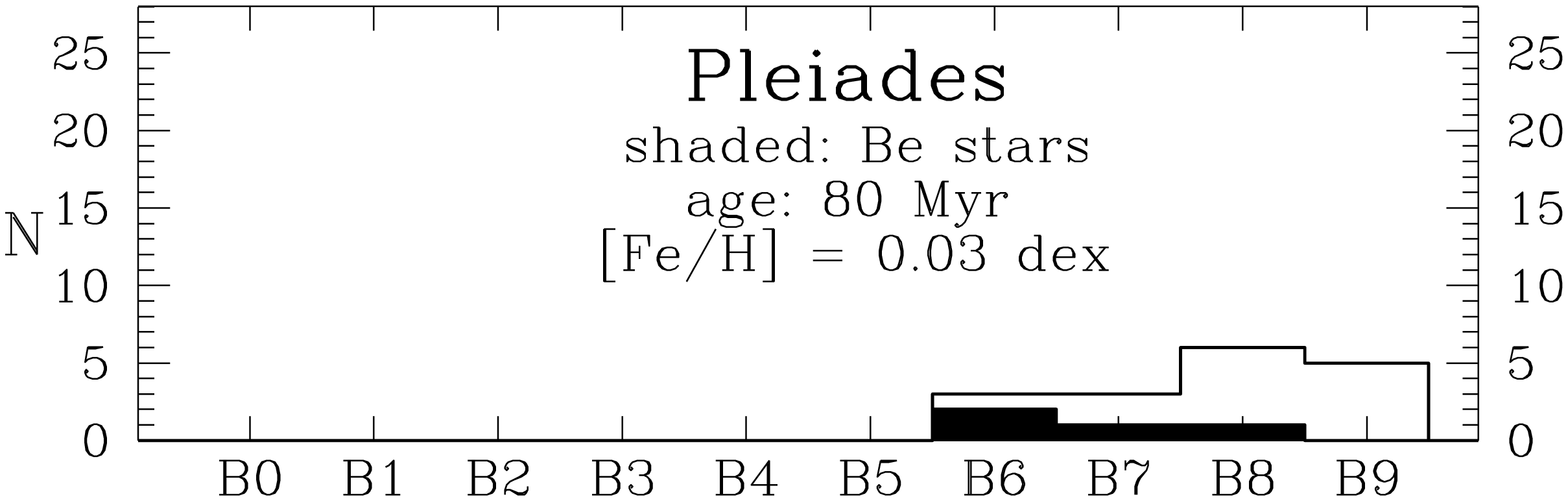,height=2.0cm,width=8.3cm,angle=-90}
}}
\centerline{\hbox{
\psfig{figure=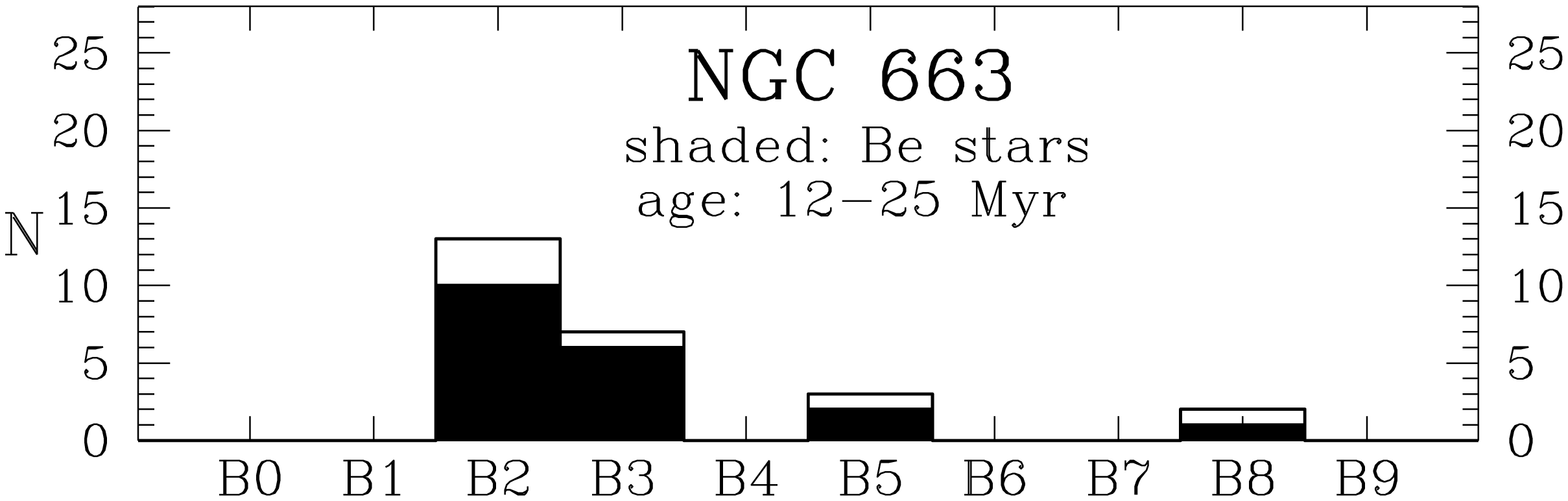,height=2.0cm,width=8.3cm,angle=-90}
\psfig{figure=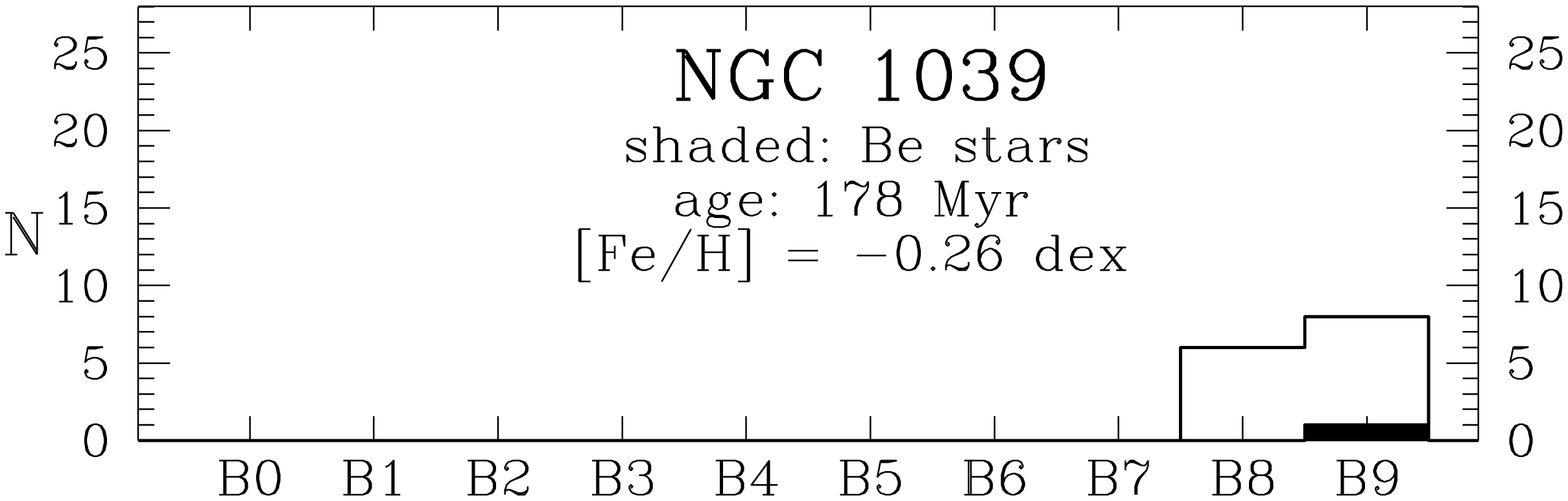,height=2.0cm,width=8.3cm,angle=-90}
}}
\centerline{\hbox{
\psfig{figure=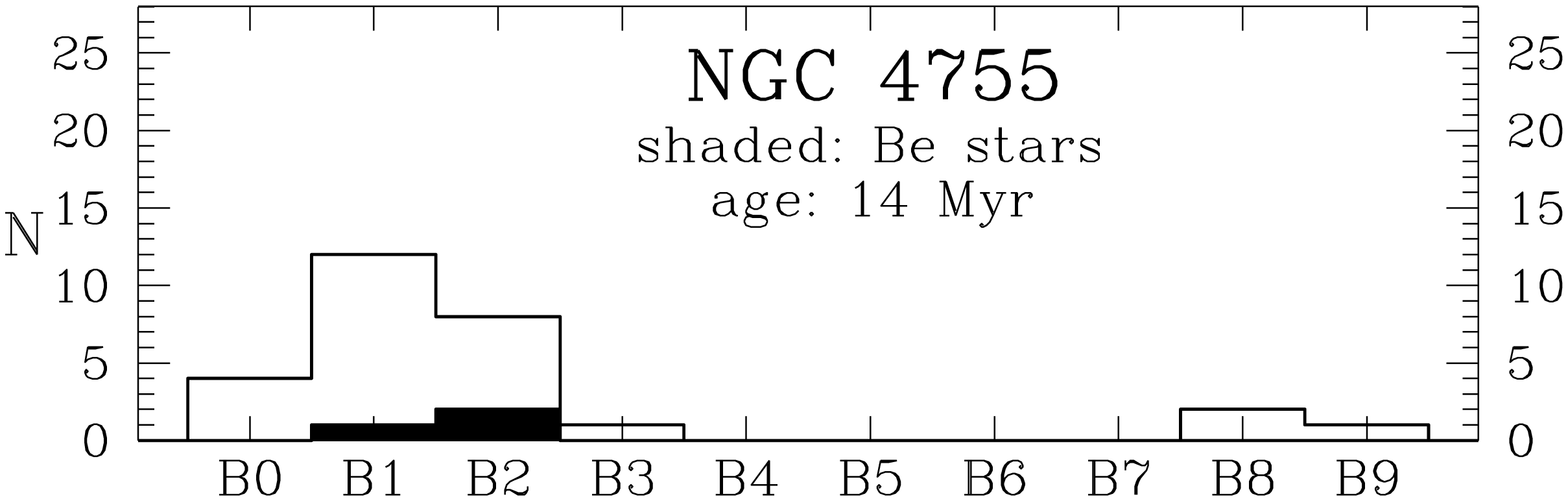,height=2.0cm,width=8.3cm,angle=-90}
\psfig{figure=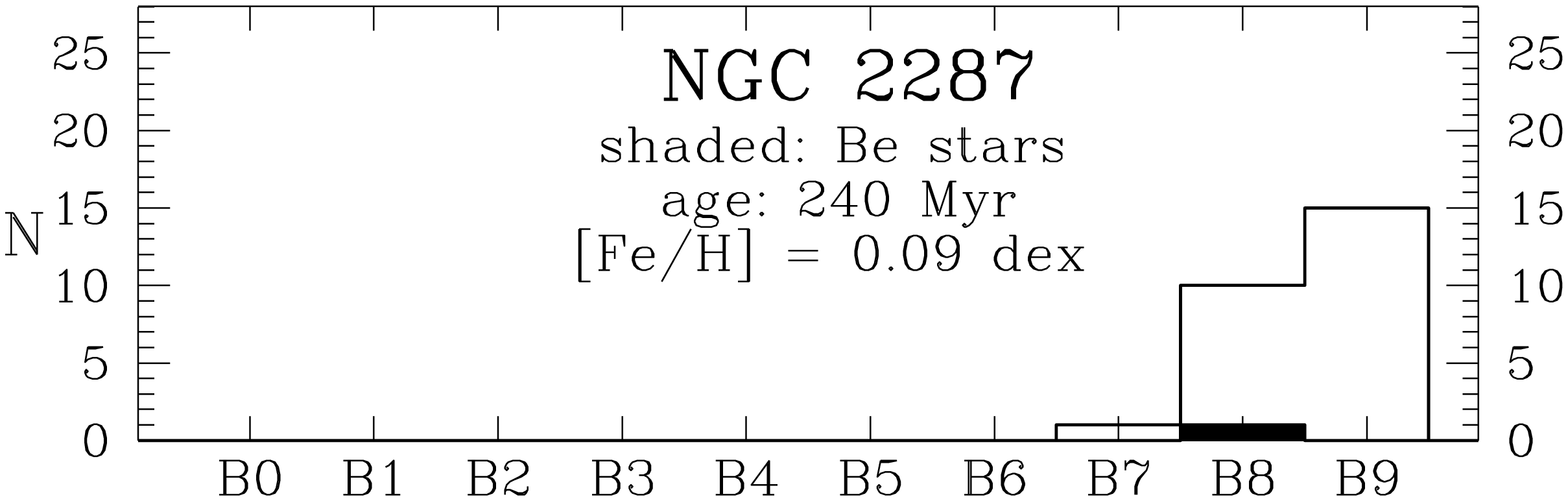,height=2.0cm,width=8.3cm,angle=-90}
}}
\caption[]{ \label{5108F5} 
Galactic Be and B stars. Only stars with spectral types and luminosity
classes III to V are plotted. ``N'' denotes total numbers of stars.
The first panel on the left shows Galactic field Be stars. The top panel
on the right presents the SMC cluster NGC 330. All other panels give 
B (white histograms) and Be stars (black histograms)
in open clusters of various ages. Iron abundances are
given when known. See Sect.\ \protect\ref{secthistphot} and 
\protect\ref{secthistspec} for references. {\em All numbers given are
only lower limits owing to incomplete availability of accurate spectral 
classifications.} This affects in particular the Be stars
among the B stars.
}
\end{figure*}

To assess the quality of the spectral typing through photometry,
we assembled a number of
Galactic open clusters for which spectroscopy of B-type stars is available.
In Fig.\ \ref{5108F5}, only stars of spectral type B {\em and}
luminosity classes ranging from V to III are considered.
It has to be emphasized that this stringent criterion introduces 
a strong selection effect not only for eligible B stars, but in particular
for Be stars. Be stars are often catalogued simply as ``Be'' without further
classification, or with spectral type but without luminosity class. 

We compiled the spectroscopic data for Galactic open clusters
from Slettebak (1985), Mermilliod 
(1988), and Buscombe (1995). The latter two catalogues are largely 
complete compilations of prior spectroscopic studies referenced therein.
We supplemented these data by photometric detections of Be stars that 
had spectroscopic classifications but had not yet previously been identified 
as Be stars. In addition to the publications referenced in Sect.\
\ref{secthistphot}, we used results from Sanduleak (1990), Goderya \&
Schmidt (1994), and Massey et al.\ (1995).  

For the clusters not already presented in Fig.\ \ref{5108F4}, we used
ages from Meynet et al.\ (1993: NGC 4755, NGC 1039, NGC 2287),
from Phelps \& Janes (1994: NGC 663), from Hillenbrand et al.\ (1993: 
NGC 6611), and else from Lyng\aa\ (1987). Metallicities were also adopted
from Lyng\aa\ as far as available. The metal abundance for $\chi$ Per was
determined by Klochkova (1991). 

Spectral classifications of B-type stars in the Magellanic Clouds exist for
many field stars and OB associations (s.\ Buscombe 1995 and references 
therein). We extracted
spectroscopic classifications for stars in 20 Lucke and Hodge (LH, 1970)
OB associations from the Buscombe catalogue. In most cases, either no Be stars 
had been detected or they were listed without MK classification. Also, in many
spectroscopic studies only the earliest B-type stars had been classified. 
These data cannot serve to determine fractions of Be stars.
Therefore, in Fig.\ \ref{5108F5} only NGC 330 is included based on the 
determination and compilation of spectral types in Grebel et al.\ (1996). 
NGC 330 is the only cluster of our set with a larger database of MK
classifications among its B-type stars. Based on their position in a 
CMD, the few Be stars classified as B2 IIe have also been included in this
diagram, since they have fainter $V$ magnitudes than the regular blue 
supergiants in NGC 330 and may be extreme Be stars as discussed by Schild
(1966) and Garmany \& Humphreys (1985).  

As explained before, field Be stars are not well suited for comparisons
to cluster Be stars because they span a range of ages and may have a range
of metallicities. On the other hand, they may give a better idea of the 
overall distribution of Be stars of different spectral types, since many
more field Be stars have spectral classifications than Be stars within a
specific cluster. Several Be star catalogues are available electronically.
We found the catalogue of Page (1984) to give the best number statistics.
From this catalogue, we extracted all non-supergiant field Be stars. The 1145
field Be stars are plotted in the first panel of Fig.\ \ref{5108F5}.
In all cases, we counted stars classified as B1.5 as B1 stars, B2.5 stars
as B2 stars, etc. Page's (1984) catalogue lists for each star all
previous spectral classifications, and different studies may differ by several
subclasses. In those cases, we usually chose the most frequently occurring
classification, while in other Be star catalogues preference has been given 
to spectral classifications from specific sources. E.g., Jaschek \&
Egret's (1982) catalogue prefers the Michigan Spectral Survey (Houk \& 
Cowley 1975, Houk 1978).

In good agreement with earlier studies, the majority of the field 
Be stars is found among the early B types. The most pronounced
peak can be seen for B2 emission-line stars (Slettebak 1982, Jaschek \&
Jaschek 1983). The B1 bin contains the second highest number of Be stars, which
is in good agreement with the Jaschek \& Egret (1982) catalogue when using the 
same spectral type binning. A small peak is seen for B8 stars in agreement
with Jaschek \& Egret (1982) and Jaschek \& Jaschek (1983). The 
small peak at B5 may be an artifact of the preferred classification of stars
as B3 or B5 rather than B4.  
 
Mermilliod (1982) pointed out that the overall Be star frequency is highest
for young clusters containing early-type stars, while total numbers decrease 
for older clusters. In Fig.\ \ref{5108F5}, open clusters are arranged to
form a sequence of increasing ages. We note in passing that age determinations
for OB clusters may vary by up to a factor of two in different studies
(e.g., Pleiades: 80 Myr according to Meynet et al.\ 1993, 150 Myr according
to Mazzei \& Pigatto 1989; $\alpha$ Per: 52 Myr following Meynet et al.\ 1993,
80 Myr according to Prosser 1992). Though small number statistics and the
above discussed incompleteness affect our diagrams, they nicely demonstrate 
Mermilliod's finding. Frequency peaks in the Be star distribution are found
at B1 to B2 and, in the case of NGC 6611, for which spectral typing for the
entire range of B stars is the most complete, at B8. As emphasized before,
all B star numbers and especially Be star numbers represent only lower limits.

Despite incomplete spectral coverage, Be star numbers vary significantly
from cluster to cluster. A young age alone does not guarantee a high number
of Be stars (compare NGC 2244 to h \& $\chi$ Per, NGC 6611 and NGC 663 to
IC 2944, NGC 4755, and NGC 2264, etc.). Neither does metallicity, which 
for the displayed Galactic clusters with known metal abundances is quite
similar. A likely second
parameter for the formation of Be stars is rapid rotation -- e.g., the stars
in h and $\chi$ Per, the young Galactic double cluster so rich in Be stars, are
known to be rapid rotators (Slettebak 1968). The much older
$\alpha$ Per, where also lower-mass stars often show H$\alpha$ emission and
Be stars start with intermediate B spectral types, contains plenty of rapid
rotators (e.g., Prosser 1992). The even older Pleiades with their late-type
Be stars also show quite high $v$ sin $i$ values (e.g., Abt 1970). 
In contrast, IC 4665 (age: 36 Myr according to Lyng\aa\ 1987) contains very
few Be stars (not plotted in Fig.\ \ref{5108F5}, but see Buscombe 1995). 
At the same time, it is known for its large fraction of binaries and, perhaps
because of that, the low rotational velocities of its members (e.g., Schild
1967).
 
The young SMC cluster NGC 330 presents a pronounced frequency peak at B2
(Fig.\ \ref{5108F5}). Almost all of the non-supergiant B-type stars 
in this cluster for which spectroscopy exists have been found to be Be
stars (Lennon et al.\ 1994, Grebel et al.\ 1996) and, as is evident 
from the diagram, most of them are B2 stars. Spectral classifications
have only been performed for the brightest stars in this cluster which 
explains the lack of later Be types. 

A comparison of the diagrams of NGC 330, NGC 3766, $\alpha$ Per, and the
Pleiades in Fig.\ \ref{5108F4} and Fig.\ \ref{5108F5} shows 
remarkable differences. Spectral typing through the photometric calibration
used in Sect.\ \ref{secthistspec} results in spectral types that are too
early by two to three spectral subclasses. A comparison of individual stars
shows, for instance, that most of the stars in NGC 330 that were assigned
types of B0 and B1 are in fact B2 stars according to classification based
on spectroscopy. For the Pleiades, the earliest B stars are of
type B6 (Johnson \& Morgan 1953), while the photometric calibration makes 
them B3 stars. This
demonstrates that even very careful photometric calibrations are
no substitute for spectroscopic classifications. It also implies that as 
of now, we cannot yet compare directly Be star fractions in young Magellanic
Cloud clusters to Galactic open clusters. A lot more spectral classification
work is needed for the Magellanic Cloud clusters, and it would be highly
desirable also to complete spectroscopic studies of the much more easily
accessible Galactic clusters.

\section{Constraints on Be star theories}\label{sectbe_constr}
 
Our current knowledge about Be stars in the Magellanic Clouds does not yet 
allow to constrain Be star theories. As was emphasized by Smith (1988),
observations of Galactic Be stars have shown that there seem to be at least 
three independent causes for the Be star phenomenon.  It appears that 
rotation, pulsation, and multiplicity may all play a role. In the case of 
NGC 330, at least some of the Be stars are members of binary systems (Grebel 
et al.\ 1996). Many Galactic Be stars, perhaps even most, are members of 
binary systems (see discussion in Abt et al.\ 1987). Pols et al.\ (1991) 
present models explaining the formation of Be stars in close binary systems. 
Pols \& Marinus (1994) use Monte Carlo simulations that convincingly reproduce 
binary main sequences as observed in Galactic open clusters (e.g., Mermilliod 
1992, Mermilliod \& Maeder 1986) and lead to the formation of blue stragglers 
and Be stars.  The number of {\em close, interacting} binaries, however, is
probably small (Slettebak 1988) both in the Milky Way and in the Magellanic 
Clouds (Van Bever \& Vanbeveren 1996).
 
That either rotation (Balona 1990, Balona et al.\ 1991) or non-radial pulsation 
(Baade 1987, 1988) plays a role in NGC 330 is shown by the presence of
$\lambda$ Eri variables discovered by Balona (1992). 
NGC 2004, which we find to be not as rich in Be
stars as NGC 330 and NGC 1818, also appears to contain only very few $\lambda$
Eri variables (Balona 1993). 
Balona (1990) suggests that the correlation between
projected rotational velocities and photometric periods of Be stars may be
explained when the photometric period equals the rotational period.
Owocki et al.\ (1994) predict that stars rotating with 90\% of $v_{\rm crit}$
should exhibit strong variability. 
High-resolution spectroscopy is needed to determine the rotational 
velocities of the variable stars in Magellanic Cloud clusters and to detect
possible signatures of non-radial pulsations, or their absence. This will
require a substantial amount of observing time at the new generation of
very large telescopes. 
 
A very elegant theory for the formation of equatorial disks, which is
capable to account for many of the observed features in Be stars was
presented by Bjorkman \& Cassinelli (1993). This wind-compressed disk
model for line-driven winds was then extended by Owocki et al.\ (1994) 
to include a full dynamical treatment of gas pressure and radiative driving.  
Disks form in the
following way for stars with large enough rotation thresholds: At low
latitudes near the surface of a star, the gravity acceleration is larger
than the radiative acceleration, and fluid streamlines ``fall'' toward the
equator, where fluids from both hemispheres collide. Standing shocks form
above and below the equator, between which a dense equatorial disk is
confined by the ram pressure of the wind.
A disk will form when the rotation rate is above a threshold value
depending on the ratio of the terminal velocity of the wind ($v_{\infty }$)
and the escape velocity from the star ($v_{\rm esc}$). For spectral types
from O to B2, this velocity ratio decreases as the terminal velocity
decreases. For instance, an O6 star has to rotate at 84\% of its critical
velocity ($v_{\rm crit}$) in order to form a disk, while for a B2 star 
48\% of $v_{\rm crit}$ suffices. Thus from the
theoretical predictions the frequency peak of Be stars is expected at B2,
which seems to be confirmed observationally (Sect.\ \ref{secthistspec}).
For a $v_{\rm crit} = 480$ km s$^{-1}$, the rotation threshold
for a B2 star is $v_{\rm rot} > 180$ km s$^{-1}$ assuming a uniform
distribution of inclination angles (sin($i$) = $\pi / 4$). Observationally,
however,  $v_{\infty }/v_{\rm esc}$ slowly continues to drop, which
Bjorkman \& Cassinelli (1993) attribute to a possible underestimation of
$v_{\infty }$ for B stars of later types because of weak line profiles of
C{\sc iv} and Si{\sc iv}.  Currently, the predicted disk
density is by a factor of 100 too small to reproduce the observed IR excess,
H$\alpha$ emission, and polarization. This could be remedied if mass-loss
rates are in fact higher than currently estimated, which reduces the
initial acceleration of the wind by decreasing the radiative line-driving
force. Furthermore, while the current model predicts both infall and outflow
from the equatorial disk, the infall could be prevented by rotationally
supporting the disk material. A weak magnetic field may already be sufficient
to add angular momentum to the disk (Bjorkman 1994).
 
The Be star phenomenon should thus be enhanced by reducing the terminal
velocity of the wind. The more metal-rich a star is, the more photons will
be absorbed, which transfers their momentum to the absorbing material and
accelerates the gas by radiation pressure. This metal-rich wind is emitted
radially and may have too high velocities to be withheld by the centripetal
force, thus gets lost and cannot contribute to forming a disk. However,
in a {\em metal-poor} environment, the radiation pressure is obviously lower,
which in turn leads to lower terminal velocities. In a metal-poor environment,
stars are more likely to have terminal velocities that are smaller than
the centripetal acceleration, thus the gas slows down and is more easily
confined. According to this intuitive scenario (Bjorkman 1995), Be stars 
should form preferably in metal-poor environments.  Lower wind speeds will 
generally lead to the formation of stronger disks (Owocki et al.\ 1994). 
However, the latest calculations by Owocki et al.\ (1996) indicate that 
the wind-compressed disk model does not work when non-radial forces and 
distortion due to rapid rotation are considered. Both lead to stronger 
poleward components of the radiative flux and prevent the formation of 
disks.

As we discussed in Sect.\ \ref{sectbe_prob}, the currently available data
do not yet allow one to make quantitative statements about Be star fractions
both in Galactic and in Magellanic Cloud clusters. Clearly, additional
spectral classification and high-resolution spectroscopic 
studies are needed. However, we can already state
that NGC 330 and NGC 1818 are rich in Be stars. High Be star fractions
are also found in more metal-rich Galactic open clusters (Sect.\
\ref{sectbe_prob}). It would be interesting to determine
terminal gas velocities for stars in these clusters. Of course, other
factors, such as intrinsic rapid rotation, will contribute as well  to the
Be star phenomenon -- perhaps
stars in the metal-poor LMC cluster NGC 2004, which has a lower Be-star
frequency, are rotating at lower velocities.

The evolutionary status of Be stars has long been under debate. 
That many of them appear spectroscopically and photometrically as
subgiants, giants, or even bright giants may be due to 
effects of rapid rotation, binarity or multiplicity,
and (infra-)red excess. The effects of rotation on spectral
classification and on the position in the CMD were studied by Slettebak et
al.\ (1980) and Collins et al.\ (1991), and Pols \& Marinus (1994) showed
the influence of binarity on the position of main-sequence stars. 
These effects can easily make main-sequence stars appear as giants.
Furthermore, Be stars are known as rapid rotators. Finally, reddening caused
by the circumstellar disks and the red excess of Be stars will move them
further on to lower temperatures. 
The study of Be stars in clusters, i.e., stellar aggregates that are supposedly
coeval and of the same origin, may help to constrain the evolutionary status
of Be stars. The young ages of our three Magellanic Cloud 
clusters make it seem likely that at least there the non-supergiant 
B-type stars are still 
burning hydrogen in their cores and thus are in the evolutionary phase of
main-sequence stars (Grebel et al.\ 1996). 

\section{Summary}

We have combined new $R$ and $H\alpha$ photometry with our $B, V$ photometry
of NGC 1818 (Will et al.\ 1995b). 
We have used the combined data set to investigate
the Be star content of NGC 1818 and surroundings using our photometric
detection method (Grebel et al.\ 1992, 1994). 
A table with photometry, astrometric positions, and designations
complying with IAU recommendations for all detected Be stars down to 18th 
magnitude in NGC 1818 and surroundings is available electronically from CDS,
Strasbourg.

Re-analyzing the NGC 1818 photometry, we find a small cluster $1\farcm5$
southeast of NGC 1818 apparently not mentioned in the literature before, 
which we have named NGC 1818 B. This cluster seems to have an age 
of at least 30 Myr and a reddening of $E(B-V)_0 = 0.11$ mag, 0.04 
mag higher than what we found for the similarly old NGC 1818. 
Differential reddening does not seem to be present across NGC 1818. 
Due to its sparse
main sequence and lack of evolved stars, it is difficult to assign an age to
NGC 1818 B. 
The current data do not allow to determine whether
the small cluster is associated with NGC 1818.  

In both clusters, we find a large number of Be stars, while the field has
almost none. The field shows a few young, massive
stars comparable or slightly older than NGC 1818 in age. The majority
of the field population appears to be of intermediate age. 

For a valid comparison with Galactic Be stars, fractions in Galactic 
clusters comparable in age have to be compared to the Magellanic clusters.
Only Be stars in clusters are likely to be coeval, equidistant, of the 
same metallicity, and to have a common origin. 
In attempting to determine Be star fractions in Galactic open clusters and
in the three Magellanic Cloud clusters that we have analyzed to date, NGC 
1818, NGC 2004, and NGC 330, we find that assigning B spectral types through 
photometry leads to systematic shifts to too early spectral types when 
compared to spectroscopic classifications. Only spectroscopy gives reliable
spectral classifications. A comparison of currently available spectroscopic
classifications for B and Be stars in clusters appears to confirm that there
is a frequency peak at types B1 to B2, as suggested by 
Mermilliod (1982) for open clusters and from the frequency distribution
of field Be stars. Quantitative determinations of
Be star fractions and studies of possible dependence of Be star numbers on
other parameters such as metallicity can only be performed once a more 
complete spectral classification coverage will be obtained for both 
Magellanic and Galactic clusters. 

On the basis of the current data, it cannot be assessed whether low
low metallicity facilitates the formation of Be stars due to lower wind 
terminal velocities.  High-resolution spectroscopy of a representative
sample of Magellanic Cloud Be stars is needed.

\section*{Acknowledgements}

During most of this work, I was supported through a Graduate Fellowship in 
the Graduiertenkolleg ``The Magellanic system, its structure, and its 
interaction with the Milky Way'' of the German Research Foundation (DFG) 
at the University of Bonn. Part of the data were obtained during my Student 
Fellowship at the European Southern Observatory, La Silla, Chile. 
Dr.\ H.E.\ Schwarz kindly obtained the $H\alpha$ frames of NGC 1818 at the NTT
for me.

I am indebted to Dr.\ You-Hua Chu for her kind support through grant NASA 
STI6122.01-94A during my stay at the University of Illinois in 
Urbana-Champaign in the final stages of this paper, and to Drs.\ H\'el\`ene 
and John Dickel for their generous hospitality. 

I am grateful to the referee, Dr.\ Arne Slettebak, for useful criticism and 
important comments, and to Drs.\ Karen \& John Bjorkman, Wolfgang Brandner, 
Katy Garmany, Stan Owocki, and Jean-Marie Will for valuable discussions.  
I thank Drs.\ 
You-Hua Chu, Klaas S.\ de Boer and Wolfgang Brandner for a critical reading
of the text. Dr.\ H\'el\`ene Dickel, chair of IAU Commission 5 Task Group on 
Designations, kindly advised me on choosing Be star designations conforming 
to IAU specifications.

This research has made use of 
the NASA Astrophysics Data System, operated at CfA, Harvard, USA, 
of the CDS catalogue service, operated at CDS, Strasbourg, France,
and of the Astronomical Data Center, operated at NASA's Goddard
Space Flight Center. I have also made use of the Digitized Sky Survey
produced at the Space Telescope Science Institute under U.S. Government
grant NAG W-2166. The southern part of the Digitized Sky Survey is based
on photographic data obtained using the UK Schmidt Telescope operated by
the Royal Observatory Edinburgh and the Anglo-Australian Observatory.

\end{document}